\documentclass[11pt,prd,superscriptaddress,preprintnumbers,amsmath,amssymb,nofootinbib]{revtex4-2}
\usepackage{epsfig}  
\usepackage{graphicx}
\usepackage{hyperref}
\usepackage{color}
\usepackage{float}
\usepackage{amsfonts}
\usepackage{amsmath}
\usepackage{slashed}
\usepackage{soul}

\large

\begin{document}

\title{UHE neutrinos encountering decaying and non-decaying magnetic fields of compact stars}
	
\author{Neetu Raj Singh Chundawat}
\email{chundawat.1@iitj.ac.in}
\affiliation{Indian Institute of Technology Jodhpur, Jodhpur 342037, India}

\author{Arindam Mandal}
\email{mandal.3@iitj.ac.in}
\affiliation{Indian Institute of Technology Jodhpur, Jodhpur 342037, India}

\author{Trisha Sarkar}
\email{sarkar.2@iitj.ac.in}
\affiliation{Indian Institute of Technology Jodhpur, Jodhpur 342037, India}

	%\date{\today} 
	
\begin{abstract}
The phenomena of neutrino spin flavour precession in the presence of an extraneous magnetic field is a repercussion of neutrino magnetic moment which is consociated with the physics beyond the standard model of electroweak interactions. Ultra high energy neutrinos are spawned from a number of sources in the universe including the highly energetic astrophysical objects such as active galactic nuclei, blazar or supermassive black holes. When such high energy neutrinos pass through any compact stellar objects like neutron stars or white dwarfs, their flux can significantly reduce due to the exorbitant magnetic field provided by these compact objects. For Dirac neutrinos, such phenomena occur due to the conversion of neutrinos to their sterile counterparts. In this work, we consider a neutron star possessing a spatially varying magnetic field  which may or may not decay with time. We find that, for the non-decaying magnetic field, the flux of high energy Dirac neutrinos becomes nearly half after passing through the neutron star. The flux is further enfeebled by $\sim 10\%$ in the presence of muons inside the neutron star. For decaying magnetic field, the flux reduction is abated by $\sim 5\%$ as compared to the temporally static magnetic field. In the case of a white dwarf, the depletion of flux is lesser as compared to the neutron stars.
\end{abstract}

\maketitle
	
\newpage
%%%%%%%%%%%%%%%%%%%%%%%%%%%%%%%%%   
\section{Introduction}
\label{intro}
%%%%%%%%%%%%%%%%%%%%%%%%%%%%%%%%%

The aberrant and largish potential of ulta high energy (UHE) neutrinos of cosmic origin to probe the structure of the universe at the leviathan scales has effectively propelled them to the omphalos of high energy physics. The fact that neutrinos interact only via weak interactions, it can perambulate colossal scales without interacting with the ambient environment which includes compact objects as well as interstellar or intergalactic medium. This may enable identification of sources of  distant UHE cosmic ray sources as these neutrinos are usually generated due to interaction of cosmic rays with the cosmic microwave or infrared backgrounds. In fact, one such identification has already been accomplished \cite{IceCube:2018cha}. In case neutrino interactions do occur, the nature of interactions can be revealed if this results in the change of flux or  flavor ratios as compared to that of the vacuum oscillations. Therefore UHE neutrinos relishes unmatched potential in unsnarling remotest demesnes of our Universe.

To  capitalize on this elephantine potentiality, a profusion of experimental facilities are either under construction or planned \cite{Ackermann:2022rqc}. These can be categorized in terms of sensitivities to energies of cosmic neutrinos. For TeV-PeV energy range, neutrinos have high opacity \cite{Hussain:2006wg,Kusenko:2001gj,Hooper:2002yq,Friess:2002cc,Anchordoqui:2002vb,Anchordoqui:2005pn}. In defiance of this, a smatter of neutrinos can still sally up to thousands of km inside the Earth before interacting \cite{Bustamante:2017xuy,IceCube:2017roe,IceCube:2020rnc}. These neutrinos can be ensnared by photonic detectors  instilled in pharaonic  voume of ice or submerged in grandiose water body. In the impending future, the northern sky will be mapped by the IceCube \cite{Halzen:2010yj,Gaisser:2014foa} whereas  KM3NeT \cite{KM3Net:2016zxf}, Baikal-GVD \cite{Baikal-GVD:2019kwy}, P-ONE \cite{P-ONE:2020ljt} will monitor the southern sky. The experiments such as IceCube-Gen2 \cite{IceCube-Gen2:2020qha}, RET-N \cite{Prohira:2019glh}, TAMBO \cite{Romero-Wolf:2020pzh} and Trinity \cite{Brown:2021lef,Brown:2021ane} are sensitive to neutrinos in the range of (1 - 100)  PeV.  If neutrino energy lies in the range of EeV, the interaction length becomes achingly diminutive \cite{Connolly:2011vc,Garcia:2020jwr,Gandhi:1995tf,Gandhi:1998ri}. In such a case, the experimental set up should be  accoutred to dredge up neutrinos interacting in the atmosphere, rock, ice or water outside of the 
detector volume \cite{Palomares-Ruiz:2005npx,Venters:2019xwi,Reno:2019jtr}. The experimental facilities such as Pierre Auger Observatory \cite{Aramo:2004pr,Nitz:2021kdx,PierreAuger:2021ccl}, ANITA I-IV  \cite{ANITA:2019wyx}, ARIANNA \cite{ARIANNA:2019scz}, ARA \cite{ARA:2019wcf,ARA:2022rwq} or
are under construction RNO-G \cite{RNO-G:2020rmc,RNO-G:2021hfx} and  PUEO \cite{PUEO:2020bnn} are bestowed with such proficiencies.  

The high energy neutrinos can beseem as an important tool to quest physics beyond the standard model (SM) of electroweak interactions. In this context, one of the  less skirred aspects of neutrinos is its electromagnetic properties. In particular, if a neutrino possesses a  finite magnetic moment due to quantum loop corrections \cite{Giunti:2014ixa,Broggini:2012df} then it may have several important hegemonies for physics of high energy cosmic neutrinos. The neutrinos would then be jaundiced by any external magnetic field, particularly a strong magnetic field. For cosmic neutrinos, even a relatively small intergalactic magnetic field of the order of $\mu$G can have sobersided ramifications. Using the current upper bound on the magnetic moment of neutrinos,   it was shown that the flux of the cosmic neutrons can be depleted by half if they peregrinate sufficiently large distances in the intergalactic field \cite{Alok:2022pdn}.

Such neutrinos can also stumble onto compact objects such as neutron stars or white dwarf stars which have large magnetic fields. The presence of such fields within compact stars would effectuate spin flip oscillations rendering depletion in the neutrino flux.  In this work, we examine this aspect. We assume that high energy muon neutrinos are emitted from a point neutrino source and after travelling some distance, it encounters a compact star with a large internal magnetic field. The compact star is assumed to be located sufficiently closer to the source that the initial flux ratio remains unchanged. It would be interesting to see whether such interactions have any observational implications or not.

The plan of our work is as follows. In section \ref{spf}, we describe the dynamics of neutrino spin flip oscillations. In section \ref{ns}, we present a brief overview of neutron star and then exhibit our results for UHE neutrinos traversing through it. In the next section,  we provide results for white dwarf star. Finally, in section \ref{con}, we furnish conclusions of our work.

%%%%%%%%%%%%%%%%%%%%%%%%%%%%%%%%%%%%%%%%%%%%%%%%%%%%%%
\section{Neutrino spin flip oscillations}
\label{spf}
%%%%%%%%%%%%%%%%%%%%%%%%%%%%%%%%%%%%%%%%%%%%%%%%%%%%%%%%

The phenomenon of neutrino oscillations are experimentally well-entrenched \cite{Bahcall:2004ut,KamLAND:2002uet,KamLAND:2004mhv,Super-Kamiokande:2004orf,MINOS:2006foh,T2K:2013ppw,T2K:2013bzi}. Such oscillations are possible only if  the neutrino flavour state is a linear superposition
of the non-degenerate mass eigenstates. It is illustrious that  a massless and chiral
neutrinos cannot have a non-zero magnetic dipole moment.  However, a massive Dirac neutrino, in general, may have a magnetic moment which emerges at the
one-loop level. In the standard electroweak theory, the Dirac neutrino magnetic moment $\mu_{\nu}$ at the leading order in $m_{\ell}^2/m_W^2$ is given as \cite{Fujikawa:1980yx}
\begin{equation}
\mu_{\nu}^{\rm SM}=\frac{3eG_F m_{\nu}}{8\sqrt{2}\pi^2}\,,
\label{eq1}
\end{equation}
where $e$ is positive, $G_F$ is Fermi constant and $m_{\nu}$ ($m_{\ell}$) is neutrino (lepton) mass. The mass of $W$ boson is denoted by $m_W$. It is ostensible that at the leading order in $m_{\ell}^2/m_W^2$, $\mu_{\nu}$ is independent of $m_{\ell}$ and the PMNS matrix. Numerically
\begin{equation}
\mu_{\nu}^{\rm SM}  \approx \frac{3.2\times 10^{-19}m_{\ell}}{1 eV} \mu_B\,. 
\end{equation}

It is apparent that the neutrino magnetic moment is directly  proportional to its mass. Therefore in order to have plenteously large value of $\mu_{\nu}$ so that it can have some observational implications, one should have sufficiently large neutrino mass. However, we have tight constraints on the mass of neutrinos \cite{KATRIN:2021uub}. Therefore, it might appear that the neutrino magnetic moment cannot be enhanced enough up to a level to have any observational impact. However, many new physics models are up to this demurral to effectuate a nonzero magnetic moment without affecting the neutrino mass. For e.g., models with scalar leptoquark \cite{Povarov:2007zz,Sanchez-Velez:2022nwm}, vector leptoquarks in the TeV mass range with  couplings both to the left and right-handed  neutrinos  \cite{Chua:1998yk}, charged scalars \cite{Voloshin:1986ty}, $R$-parity breaking SUSY \cite{Aboubrahim:2013yfa, Fukuyama:2003uz}, an additional approximate horizontal symmetry $SU(2)_H$ with the SM gauge group \cite{Babu:2020ivd}, non standard interaction (NSI) \cite{Giunti:2014ixa,Healey:2013vka,Papoulias:2015iga,Kharlanov:2020cti}  can generate a non-zero neutrino magnetic moment that may lie within the experimental reach.

On the experimental front, various reactor and accelerator experiments have provided bounds on the neutrino magnetic moment. The reactor experiments include the GEMMA experiment at the Kalinin Nuclear Power Plant, Russia  \cite{Beda:2009kx} and the TEXONO experiment  at the Kuo-Sheng Reactor Neutrino Laboratory, Taiwan \cite{TEXONO:2006xds} and ROVNO  experiment at Rovno nuclear power plant, Ukraine \cite{Derbin:1993wy}. The accelerator experimental set ups entail LAPMF \cite{Allen:1992qe} and LSND \cite{LSND:2001akn}. The bounds from these reactor and accelerator neutrinos are in  the range $(10^{-11}-10^{-10})~\mu_B$.  Recently, the XENON1T experiment provided upper bound on 
 $\mu_{e\nu}$ as $(1.65-3.42)\times10^{-11}~\mu_B$ \cite{Babu:2020ivd}.
 
The limits on $\mu_{\nu}$ can also be obtained from astrophysics and cosmology through the evolution of stars \cite{Heger:2008er}, plasmon decay in the stellar environment \cite{Borisov:2014cqa}, supernova events \cite{deGouvea:2012hg} and $^4_2 He$ nucleosynthesis \cite{Vassh:2015yza}. In fact, the most stringent limits on the neutrino magnetic moment comes from the red giant branch (RGB) of globular cluster which is $\mu_{\nu}<4.5\times 10^{-12}~\mu_B$ \cite{Viaux:2013hca}. This value is  about one order of magnitude smaller than that obtained from several terrene experiments .

Since neutrinos are deprived of electric charge, they can partake in electromagnetic interactions only by coupling with photons
through quantum corrections. The Hamiltonian for the neutrino-photon interaction is given by \cite{Giunti:2014ixa}
\begin{eqnarray}\label{eq3}
\mathcal{H}_{EM}=\bar{\nu}(x)\Lambda^{\mu}\nu(x) A_{\mu}			\sim J_{\mu}^{EM}A_{\mu},
\end{eqnarray}
where $A_{\mu}$ is the electromagnetic field and $\Lambda_{\mu}$ represents the vertex function subsuming the electromagnetic properties of neutrino which is a $4\times4$ matrix in the spinor space and 
 containes all the form factors corresponding to the four electromagnetic properties, given by (a) electric moment, (b) magnetic moment, (c) neutrino charge and (d) anapole moment. In Dirac picture, the form factor corresponding to electric moment vanishes due to the assumption of $CP$ invariance and the hermiticity of the electromagnetic current, $J_{\mu}^{EM}$. The hermiticty of the current, $J_{\mu}^{EM}$ also restricts the form factors corresponding to the magnetic moment, neutrino charge and anapole moment to be real.

The magnetic moment spawns the neutrino spin to precess in the contiguity of an external  magnetic field which ushers to the mixing between the left handed and right handed neutrino current. In the two flavour framework, the Hamiltonian for the neutrino state evolution is expressed as $4\times4$ Hermitian matrix. In our work we consider only Dirac neutrinos, for which the basis vector is $(\nu_{eL},\nu_{\mu L},\nu_{eR},\nu_{\mu R})$ and the Hamiltonian in this case manifests as \cite{Giunti:2014ixa,Broggini:2012df}
\begin{equation}\label{eq6}
H_D=\begin{pmatrix}
\frac{-\Delta m^{2}}{4E_{\nu}}cos 2\theta+V_e & \frac{\Delta m^{2}}{4E_{\nu}}sin 2\theta & \mu_{ee}B_\perp & \mu_{e\mu}B_\perp\\
\frac{\Delta m^{2}}{4E_{\nu}}sin 2\theta & \frac{\Delta m^{2}}{4E_{\nu}}cos 2\theta+V_\mu & \mu_{e\mu}B_\perp & \mu_{\mu\mu}B_\perp \\
\mu^{*}_{ee}B_\perp & \mu^{*}_{\mu e}B_\perp & \frac{-\Delta m^{2}}{4E_{\nu}}cos 2\theta & \frac{\Delta m^{2}}{4E_{\nu}}sin 2\theta \\
\mu^{*}_{e\mu}B_\perp & \mu^{*}_{\mu \mu}B_\perp & \frac{\Delta m^{2}}{4E_{\nu}}sin 2\theta  &  \frac{\Delta m^{2}}{4E_{\nu}}cos 2\theta \\ 
\end{pmatrix}.
\end{equation} 
A spin rotation from negative helicity state towards positive state will
reduce the resultant effective weak neutral and
charged-current scattering cross sections for neutrino
which emerge, in the relativistic case, predominantly from the negative helicity state for neutrinos. 

In equation \eqref{eq6}, $B_\perp$ denote transverse magnetic field, while $\mu$ is the neutrino magnetic moment. Here, $\Delta m^{2}=m_{2}^2-m_{1}^2$. $V_{e}$ and $V_{\mu}$ are the potentials experienced by $\nu_e$ and $\nu_{\mu}$, respectively which are dependent on the matter composition of the medium through which the neutrino travels. If the matter is consisted of only nucleons and electron ($npe$), the matter potentials take the form of, $V_e=\sqrt{2}G_F (n_e-n_n/2)$ and $V_\mu=-\sqrt{2}G_F n_n/2$, while if muons are added with it ($npe\mu$), $V_{\mu}$ is altered to be, $V_\mu=-\sqrt{2}G_F (n_{\mu}-n_n/2)$ due to the occurence of $\nu_{\mu}$ charged current (CC) interaction in matter in addition to that of the electrons also. Here, $n_e$, $n_n$ and $n_{\mu}$ are the number densities of electron, neutron and muon respectively.

%%%%%%%%%%%%%%%%%%%%%%%%%%
\section{Cosmic neutrinos passing through neutron star}\label{ns}
%%%%%%%%%%%%%%%%%%%%%%%%%%
Neutron stars (NS) are unique compact stars engendered when the core nuclear fuel of a main sequence star of intermediate mass ($M\geq 8M_{\odot}$) is completely exhausted with the formation of the most stable iron nuclei. At this stage, the hydrodynamic equilibrium in the stellar structure is sustained by the degeneracy pressure of excess neutrons, against the inward gravitational collapse, which are the byproducts of photodisintegration and the following process of neutron drip from the iron nuclei. With further infalling matter from the outer layer towards the center, the density of the system may even exceed nuclear saturation density and the system detonates with type-II supernova (SN) explosion. As a consequence, NSs are born \cite{Lattimer:2004pg}. The typical mass and radius of a NS are of the order of $\sim 2M_{\odot}$ and $\sim 10$ km respectively, with its central density being a few times of the nuclear saturation density. The exact estimation of these quantities are obtained from its matter configuration and the equation of state (EOS). The simplest NS models consist of matter containing neutrons, protons and electrons ($npe$). However, muons start to appear in the matter when the chemical potential of the electrons exceeds the rest mass of muons ($105.6$ MeV) \cite{Zhang:2020wov}. NSs also possess extremely high magnetic fields, also known as magnetars, which can even reach as high values as $10^{18}$ G at its center \cite{Esposito:2018gvp}. The central and surface values of magnetic fields may differ by one or two orders of magnitude.

NSs are the natural sources of neutrinos having energy in the $\sim$MeV ($10^6$ eV) range which are initially trapped for a few seconds during the birth of the star. Later during the different stages of its thermal evolution, neutrinos also continue to be emitted from the star as they are the most efficient heat carriers and help to reduce the temperature of the star. These $\nu_e$ neutrinos can also undergo  spin flip precession, see for e.g.,  \cite{Adhikary:2022phm,Alok:2022ovy}.
 However, in this work we are not considering the neutrinos which are generated inside the NS. Rather, we consider UHE neutrinos generated from existing sources nearby the NS \cite{Meszaros:2017fcs}, such as active galactic nuclei (AGN) \cite{Murase:2015ndr, Schuster:2001xi} or a blazar \cite{Giommi:2021bar, Oikonomou:2019pmg}. These objects are able to produce mainly $\nu_{\mu L}$ {\footnote{For convenience, from now on, we write $\nu_{\mu}$ for $\nu_{\mu L}$.}} with energies which can transcend the $\sim$EeV ($10^{19}$ eV) range. We contemplate these UHE neutrinos peregrinating through the NS which are expected to be affected by the star's magnetic field and hence exhibit the reduction in their flux due to the SFP phenomena.  

In our analysis, we consider a massive NS dovetailed of normal matter and having mass $2.3M_{\odot}$ and radius $11.8$ km, i.e., we consider the composition of $npe$ and $npe\mu$ with large magnetic field at its core $10^{18}$ G. The matter configuration of the star is described by a nonlinear Walecka model containing $\sigma$, $\omega$ and $\rho$-mesons \cite{PhysRevC.69.045803, Mueller:1996pm}, and is governed by the GM1 equation of state (EOS) \cite{Glendenning:1991es}. The neutrinos are assumed to be Dirac in nature and have a finite magnetic moment of $\sim 10^{-11}~\mu_B$. 

In this analysis, we consider the simplification of zero vacuum mixing $(\theta_{12}=0)$ which yields the Hamiltonian to have a form of $2\times2$ matrix, given by \cite{Joshi:2019dcj}
\begin{equation}\label{eq8}
H=\begin{pmatrix}
-\Delta m_{12}^2/4E+\Delta V/2 & \mu_{e\mu}B \\
\mu_{e\mu}B & \Delta m_{12}^2/4E-\Delta V/2
\end{pmatrix}.
\end{equation}
Here $\Delta V=\sqrt{2}G_F \rho Y_e^{eff} /m_N$ with $\rho$ being the matter density. $Y_e$ and $m_n$ denote the effective electron fraction and nucleonic mass, respectively. Further, $Y_e^{eff}=(3Y_e-1)/2$ for Dirac basis. Following eqn. \eqref{eq8}, the neutrino spin flavour evolution equation in the presence of an external magnetic field is expressed as \cite{Joshi:2019dcj}
\begin{widetext}
\begin{equation}\label{eq9}
\frac{d^2 \nu_{\mu L}}{dr^2}-\left(\frac{\mu \dot{B}}{\mu B}+i\zeta \right)\frac{d\nu_{ \mu L}}{dr}+\left[\phi^2+i\frac{d\phi}{dr}+(\mu B)^2-i\phi \frac{\mu \dot{B}}{\mu B}+\phi \zeta\right]\nu_{\mu L}=0,
\end{equation}
\end{widetext}
where 
\begin{eqnarray}\label{eq10}
\phi&=&\frac{-\Delta m^2}{4E}+\frac{1}{\sqrt{2}}G_F n_e,  \nonumber\\
\zeta&=& \frac{1}{\sqrt{2}}G_F n_n,~ \nu_{\mu L}\rightarrow \nu_{\mu R}\,.
\end{eqnarray}
The transition probability of left handed electron neutrino can be obtained after solving \eqref{eq9} along with \eqref{eq10}. 

%%%%%%%%%%%%%%%%%%%%%%%%%%
\subsection{Non-decaying magnetic field}
%%%%%%%%%%%%%%%%%%%%%%%%%%
We consider the following universal profile of the internal magnetic field  of the NS
\begin{equation}\label{eqa}
B(x)=B_c(1-1.6x^2-x^4+4.2x^6-2.4x^8),
\end{equation}
where $x=r/R$, with $r$ is the radial distance from the center of the star and $R$ is the radius of the star. $B_c$ is the magnetic field at the center which is assumed to be $B_c=10^{18}$ G. The profile is independent of any free parameter which works at its advantage. The magnetic field at the surface of the star is obtained to be $\sim 10^{17}$ G, following the profile given in eqn. \eqref{eqa}.

\begin{figure}[h!]
\includegraphics[scale=0.4]{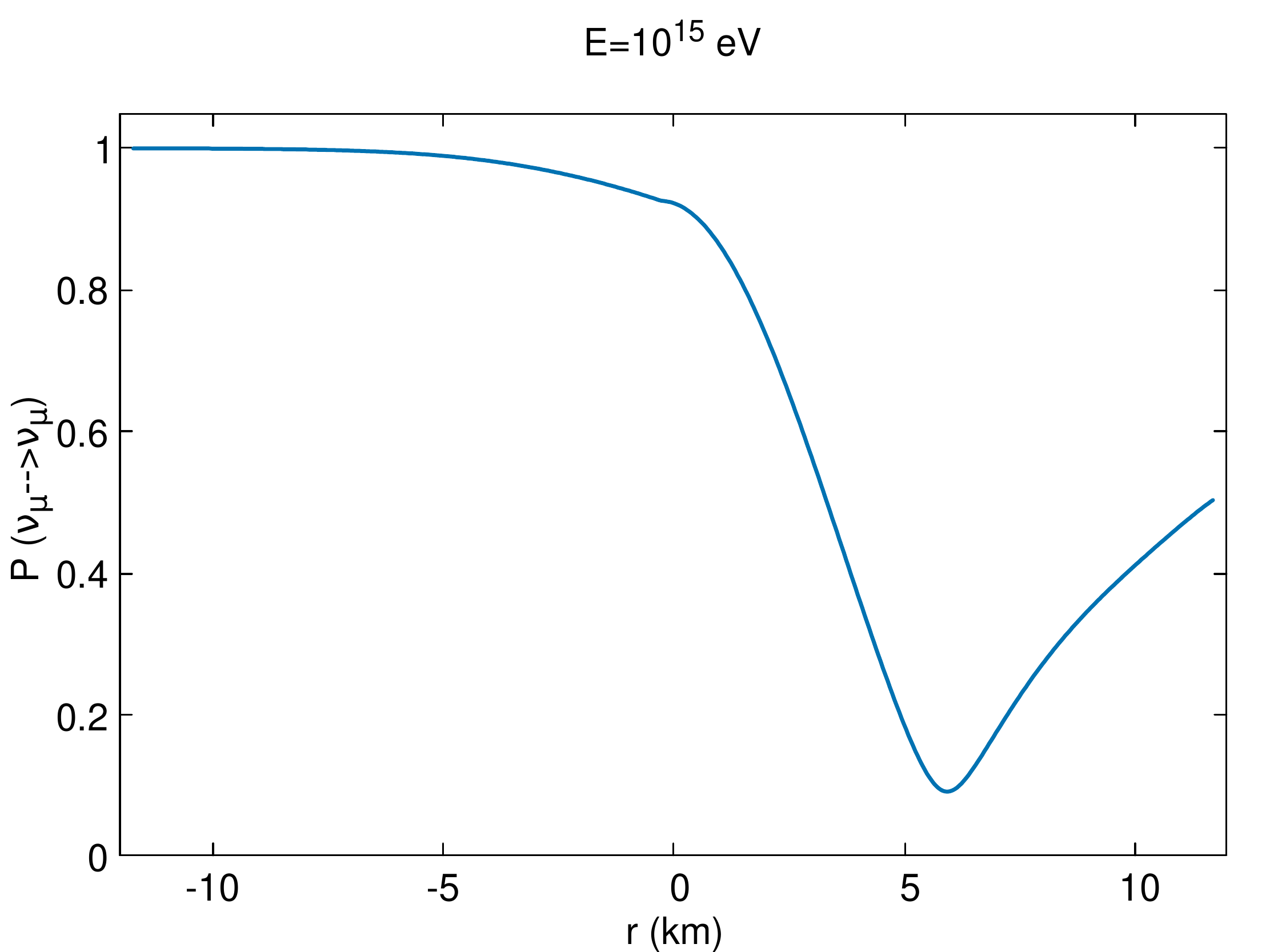}\hspace*{1mm}\includegraphics[scale=0.4]{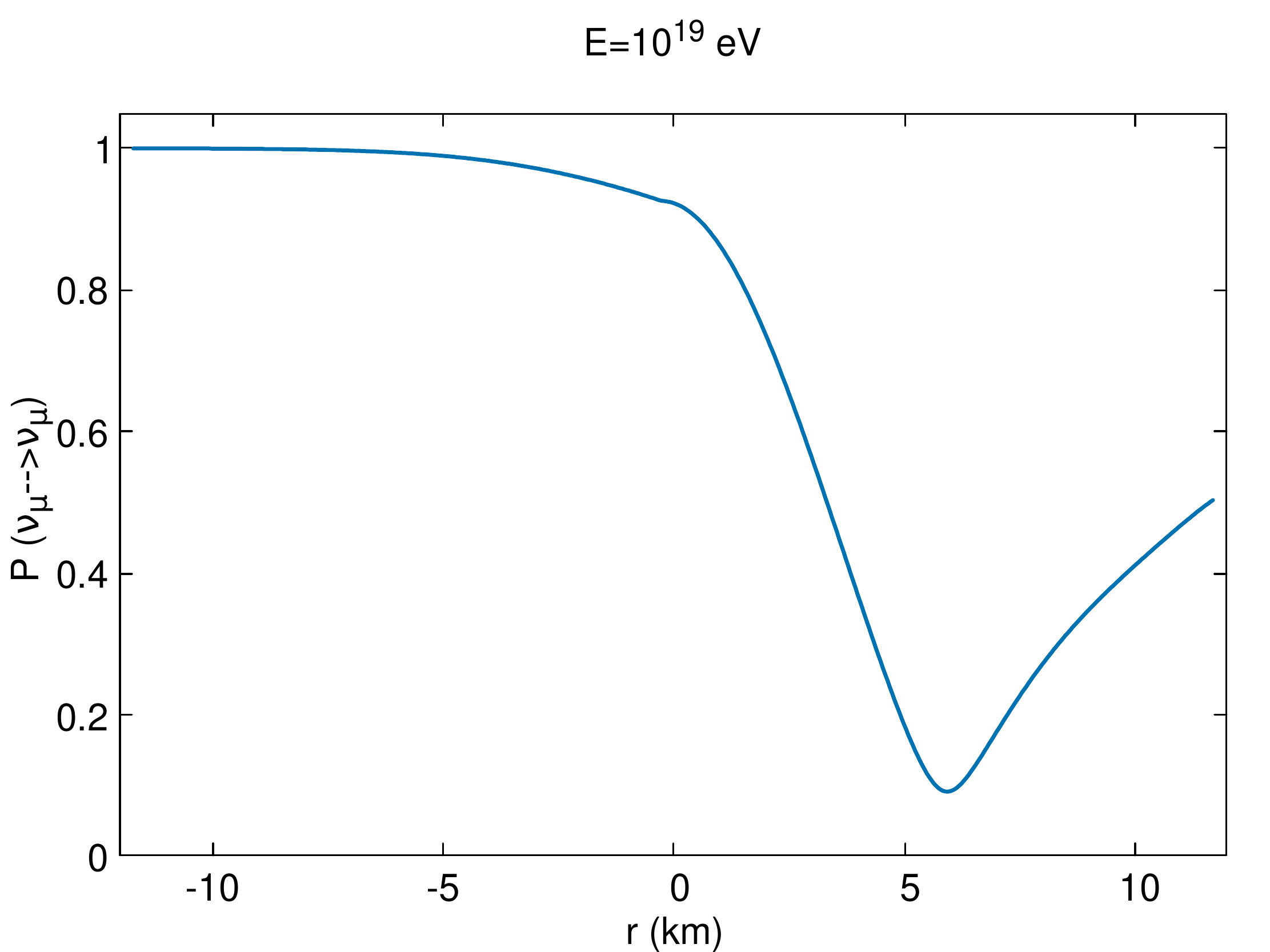}
\caption{Variation of survival probability $P(\nu_{\mu}\rightarrow \nu_{\mu})$ with the radial distance inside the star for two different energy sets of incoming $\nu_{\mu}$. The star contains nuclear matter with $npe$ configuration.}
\label{f1}
\end{figure}

Fig. \ref{f1} demystifies the survival probability of the incoming $\nu_{\mu}$ beam. In the left and right panels of the figure, we consider the neutrino energies to be $E=10^{15}$ eV and $E=10^{19}$ eV, respectively. We contemplate the magnetar as a sphere in which the neutrino beam enters at the point $r=-R$ and leaves the star from the point $r=R$. As stated before, $R$ is the radius of the star with value $11.8$ km.
We assume that the incoming UHE neutrinos travel along the diameter of the star  as being the largest traversed distance, the effect of SFP will be maximal.
It can be observed from the figure that the survival probability of $\nu_{\mu}$ at its point of emission from the star at $r=R$ is $\sim 0.5$ which implies that almost half of the incoming neutrinos are converted into their sterile counterparts. The result does not vary significantly with increase in incoming neutrino energy, as observed comparing left and right panels of the figure. The survival probability of $\nu_{\mu}$ is obviously unity at the point of entry to the star, $r=-R$, as we have considered zero vacuum mixing and hence the neutrino beam incident on the magnetar only consists of $\nu_{\mu}$.    

\begin{figure}[h!]
\includegraphics[scale=0.4]{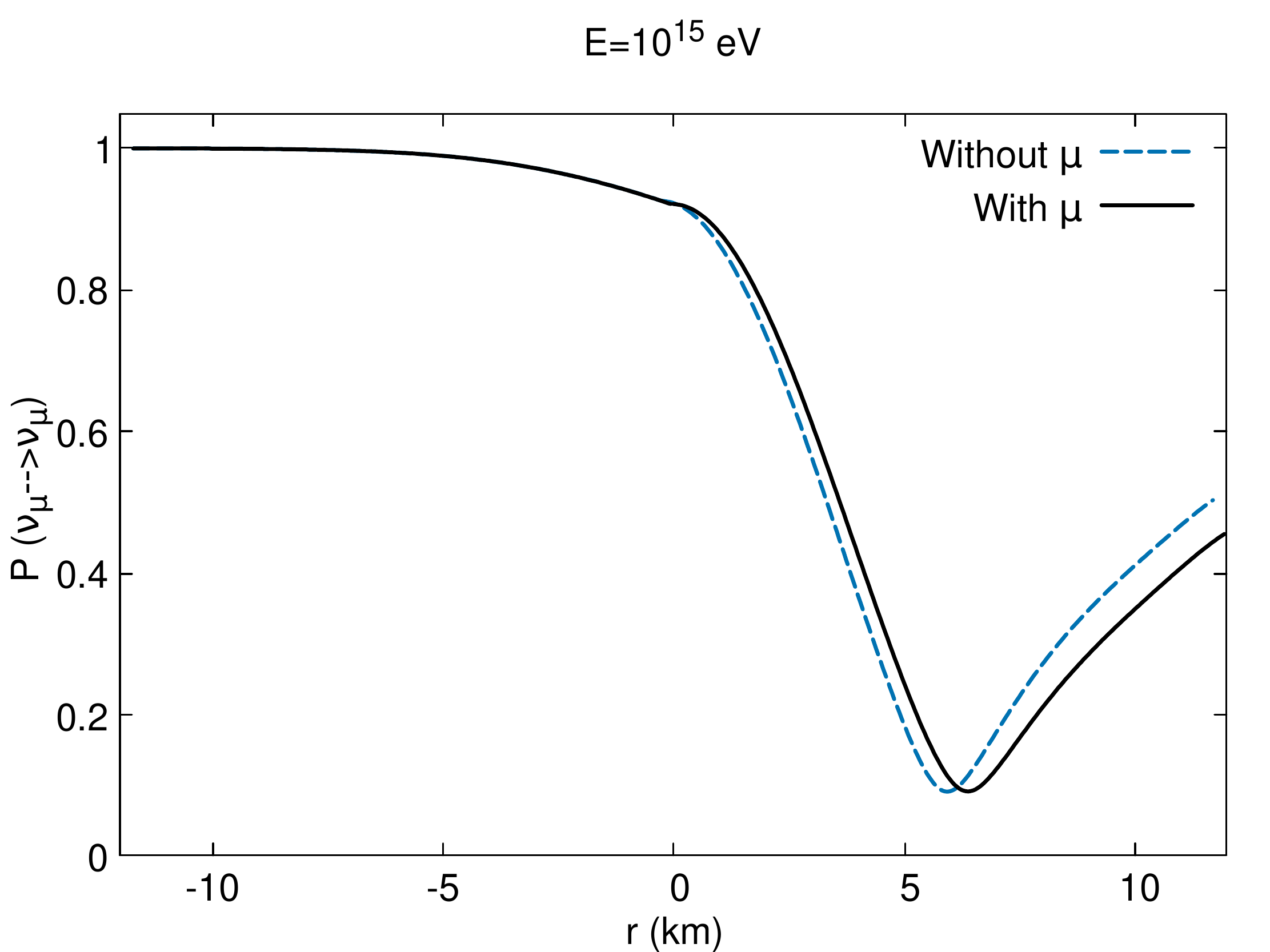}\hspace*{1mm}\includegraphics[scale=0.4]{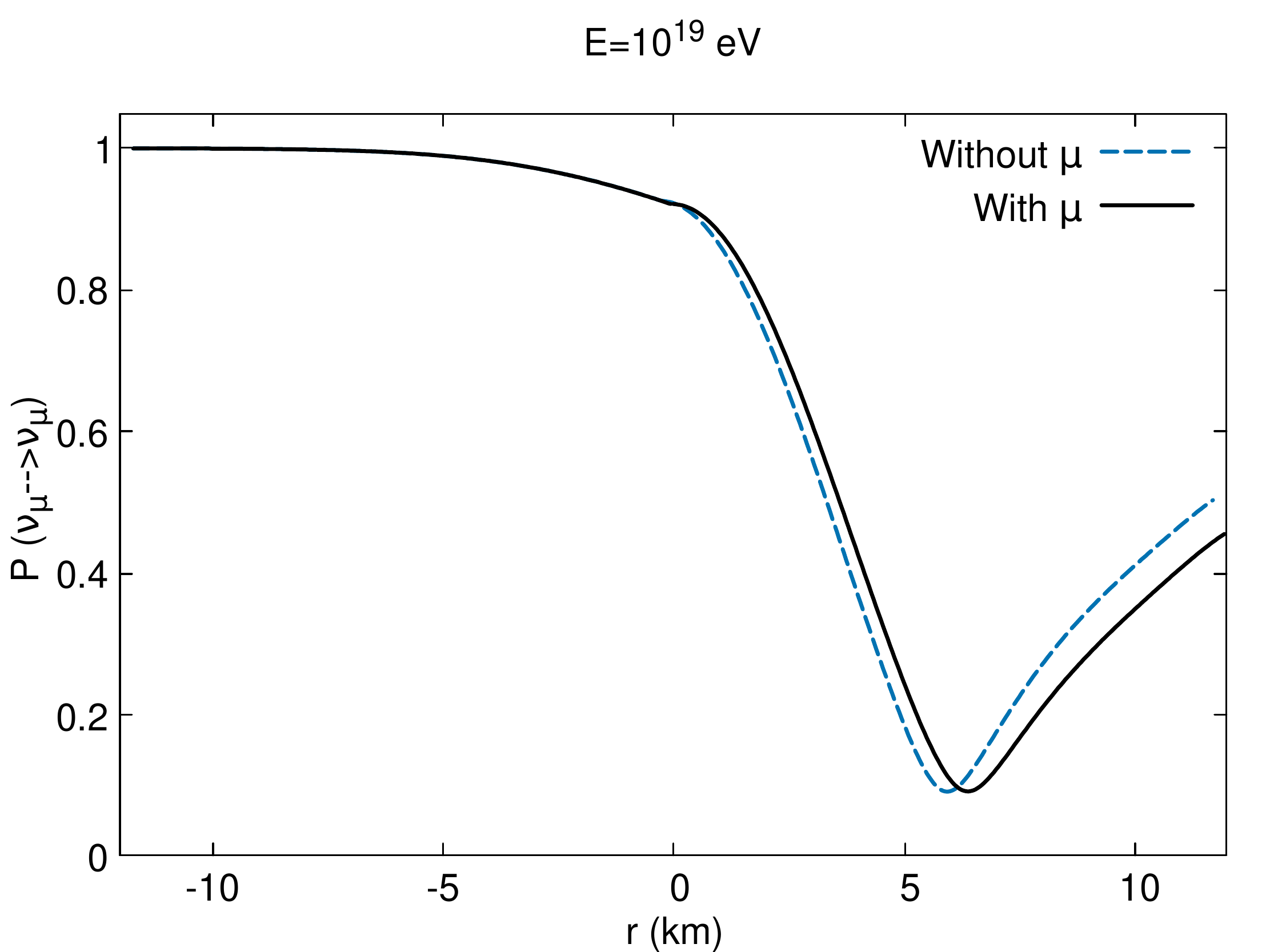}
\caption{Same as fig. 1, comparing $npe\mu$ and $npe$ matter.}
\label{f2}
\end{figure}

In fig. \ref{f2}, we portray the plot for $\nu_{\mu}$ survival probability after crossing the NS for two kinds of matter composition, $npe$ and $npe\mu$. In the left and right panels of the plot, the incoming energy of $\nu_{\mu}$ beam is considered to be $10^{15}$ eV and $10^{19}$ eV, respectively. The plots clearly show that starting from the point of incidence ($r=-R$) till the center of the star, the survival probability is nearly similar for the two kinds of matter configuration. However, the difference starts to be pronounced once it crosses the stellar center and becomes greater near the surface region. It is to be noted that in the presence of muon in NS matter, at the point of emission ($r=R$), the survival probability of $\nu_{\mu}$ is further lowered to a value of $\sim 0.45$ from $\sim 0.5$ which is the muon neutrino survival probability for only $npe$ matter at $r=R$. This implies that the presence of muon affects the SFP phenomena to a greater extent and thus aids to generate a larger flux of sterile neutrino. 
 The flux of incoming $\nu_{\mu}$ is diminished by $50\%$ after they drift through the $npe$ matter of NS whereas the reduction is enhanced to $55\%$ in the presence of muons.  The result does not exhibit any significant variation for the case of lower and higher energies as observed from assimilating the two panels of the figure. The results are summarized  in table \ref{t}.
\begin{center}
\begin{table}
  \begin{tabular}{|l|l|l|l|}
    \hline
    Conditions &
      \multicolumn{2}{c|}{$P(\nu_{\mu}\rightarrow \nu_{\mu})$ at $r=R$} &  $\%$ decrease in flux at $r=R$
       \\
       \cline{2-3}
    & $10^{15}$ eV & $10^{19}$ eV & 
     \\
     \hline
    Without muon & $0.5$ & $0.5$  &  $50\%$ 
       \\
       \hline
     With muon & $0.45$ & $0.45$  &  $55\%$
     \\   
       \hline
  \end{tabular} 
  \caption{Survival probability and the   $\nu_{\mu}$ flux reduction  at the point of emission from the star, i.e., at $r=R$, $R=11.8$ km.}
  \label{t}
\end{table}
\end{center}

%%%%%%%%%%%%%%%%%%%%%%%%%%
\subsection{Decaying magnetic field}
%%%%%%%%%%%%%%%%%%%%%%%%%%
In NS, the magnetic field starts to deplete over time which may undergo via three mechanisms: \cite{1992ApJ...395..250G} (a) Ohmic decay, in case of which the magnetic field diffuses with respect to the charged particles \cite{Sengupta:1998yk}, (b) ambipolar diffusion, which is a dissipative effect where the diffusion of electrons and protons in opposite directions in a neutron background takes place \cite{2011MNRAS.413.2021G} and (c) Hall drift \cite{2012MNRAS.421.2722K}. In our analysis, we consider the dissipation of magnetic field by the process of ambipolar diffusion which is the most important mechanism occurring inside the core of the manetar and affects the magnetic field  most significantly at the lower core temperature ($\sim 10^8$ K) \cite{2011MNRAS.413.2021G}. To incorporate the effect of magnetic field dissipation in the magnetar, we consider only the occurrence of ambipolar diffusion for the transport of magnetic flux and follow the formalism mentioned in the ref. \cite{Passamonti:2016nmf,2011MNRAS.413.2021G}, according to which the expression for time varying magnetic field in $npe$ matter, is given by
\begin{equation}
B(r,t)=\frac{B_0(r)}{1+t/\tau}\,.
\end{equation}
Here $B_0$ is the constant magnetic field without taking the decaying effect into account and is obtained from eqn. \eqref{eqa}. Further, $\tau$ is the characteristic decay time for the magnetic field due to the occurrence of ambipolar diffusion and can be expressed as, 
\begin{equation}
\tau \approx 25 \,L_5^2\, B_{17}^{-2}\, T_9^2\,(\rho/\rho_0)^{2/3}\,\, {\rm yrs},
\end{equation}
 where $L_5=L/10^5$ cm, $L$ being the distance over which the variation of magnetic field is considered, $B_{17}=B/10^{17}$ G and $T_9=T/10^{9}$ K. $\rho_0$ is the saturation nuclear density which is dependent on the choice of EOS of NS matter. For GM1 EOS, $\rho_0=0.153$ fm$^{-3}$ \cite{Watanabe:2022cuv}.

\begin{figure}[h!]
\includegraphics[scale=0.4]{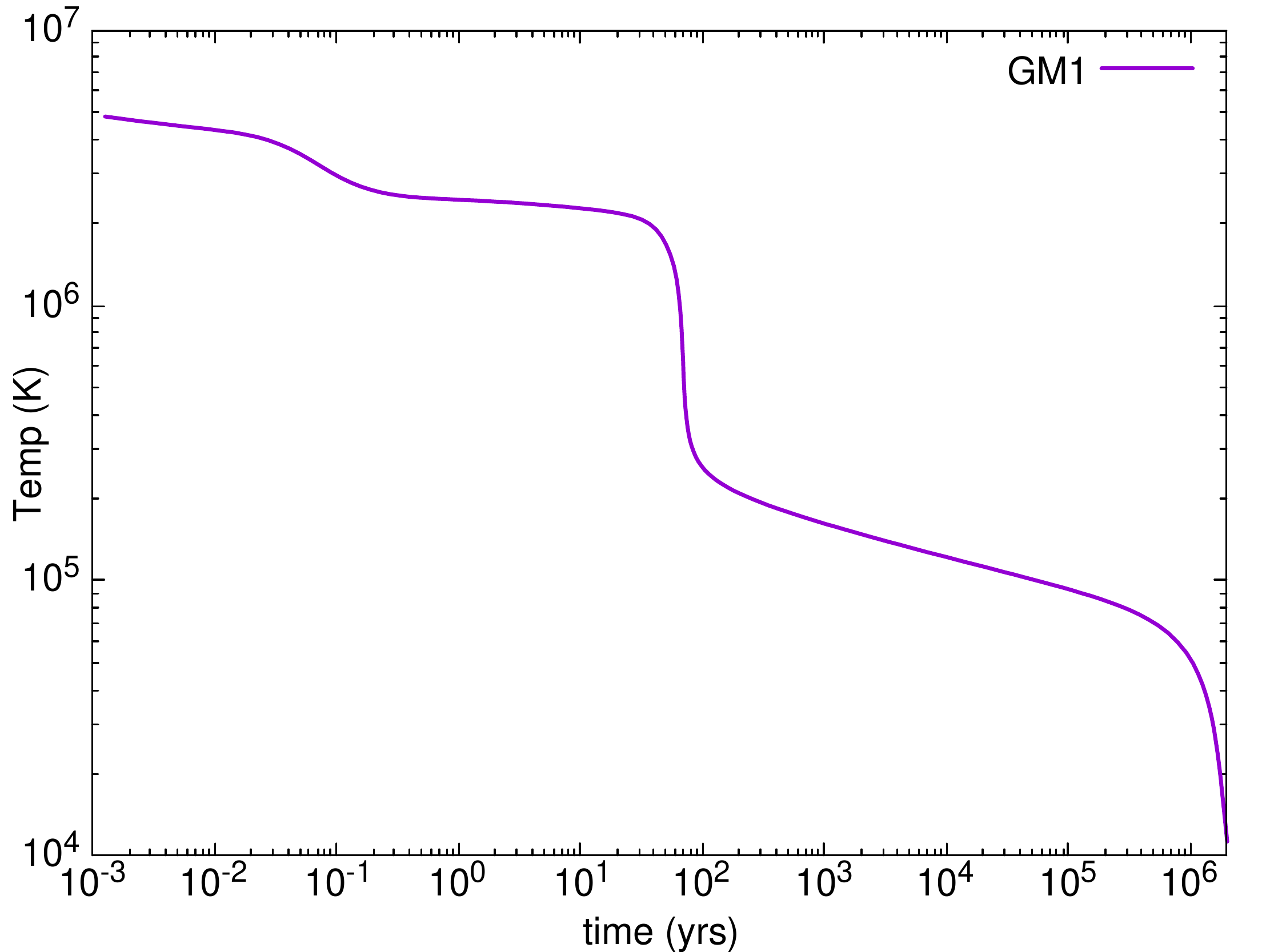}\hspace*{1mm}\includegraphics[scale=0.4]{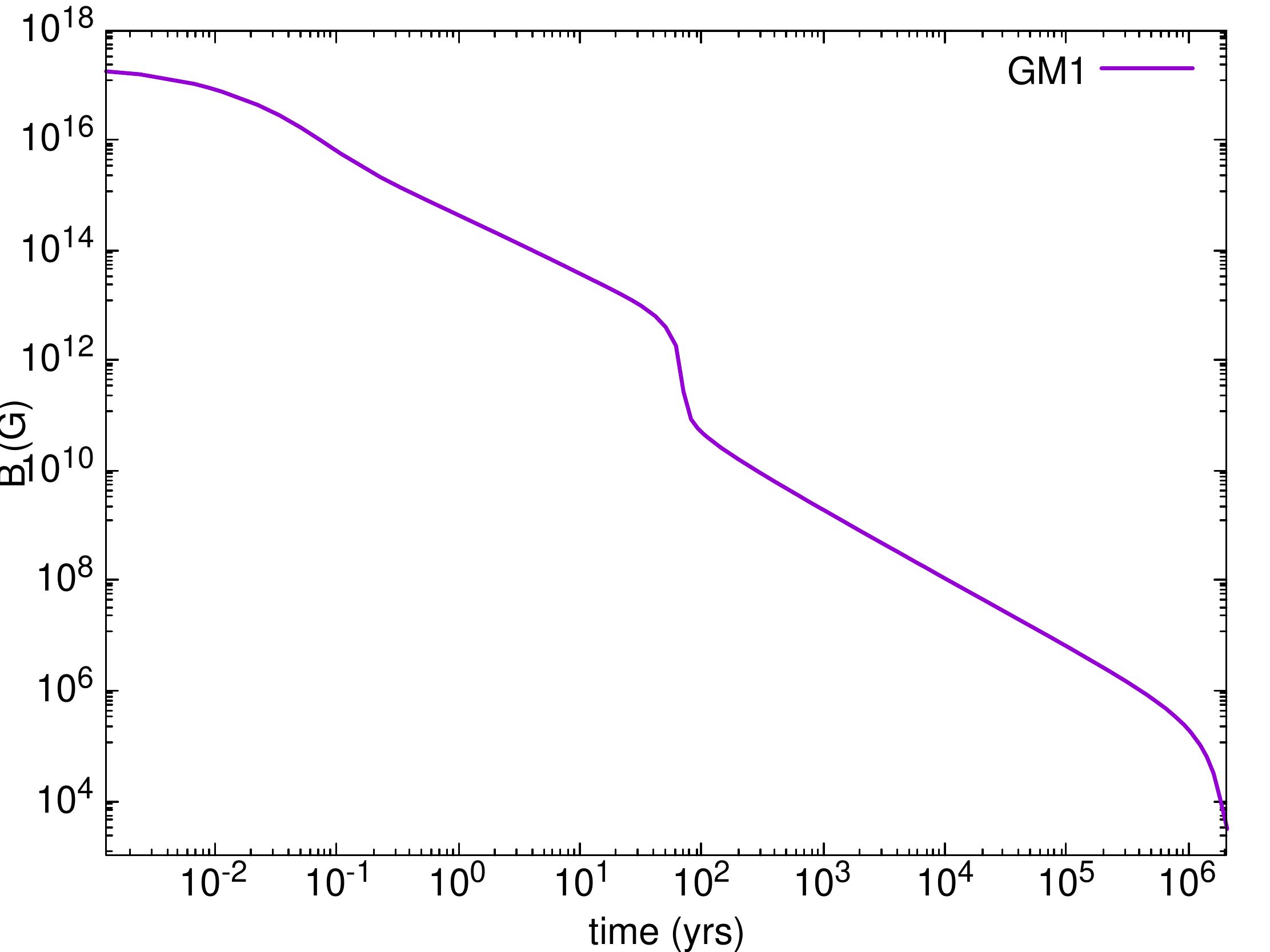}
\caption{Variation of surface temperature (left panel) and magnetic field (right panel) with time of a magnetar of mass $2.3~M_{\odot}$ and radius $11.8$ km.}
\label{ft}
\end{figure}

The surface temperature of our model star varies within a range of $\sim 10^7-10^4$ K over a temporal span of $10^6$ years, starting after the birth of the NS and the corresponding surface magnetic field varies from $\sim 10^{18}$ G to $10^{4}$ G which are depicted in the left and right panels of fig. \ref{ft}, respectively. For a massive NS like our model star, the initial thermal relaxation period of the stellar thermal evolution lasts upto $\sim 100$ years, during which the surface magnetic field also shows drastic reduction in its value from $\sim 10^{14}$ G to $10^{11}$ G, as observed from fig. \ref{ft}.

\begin{figure}[h!]
\includegraphics[scale=0.4]{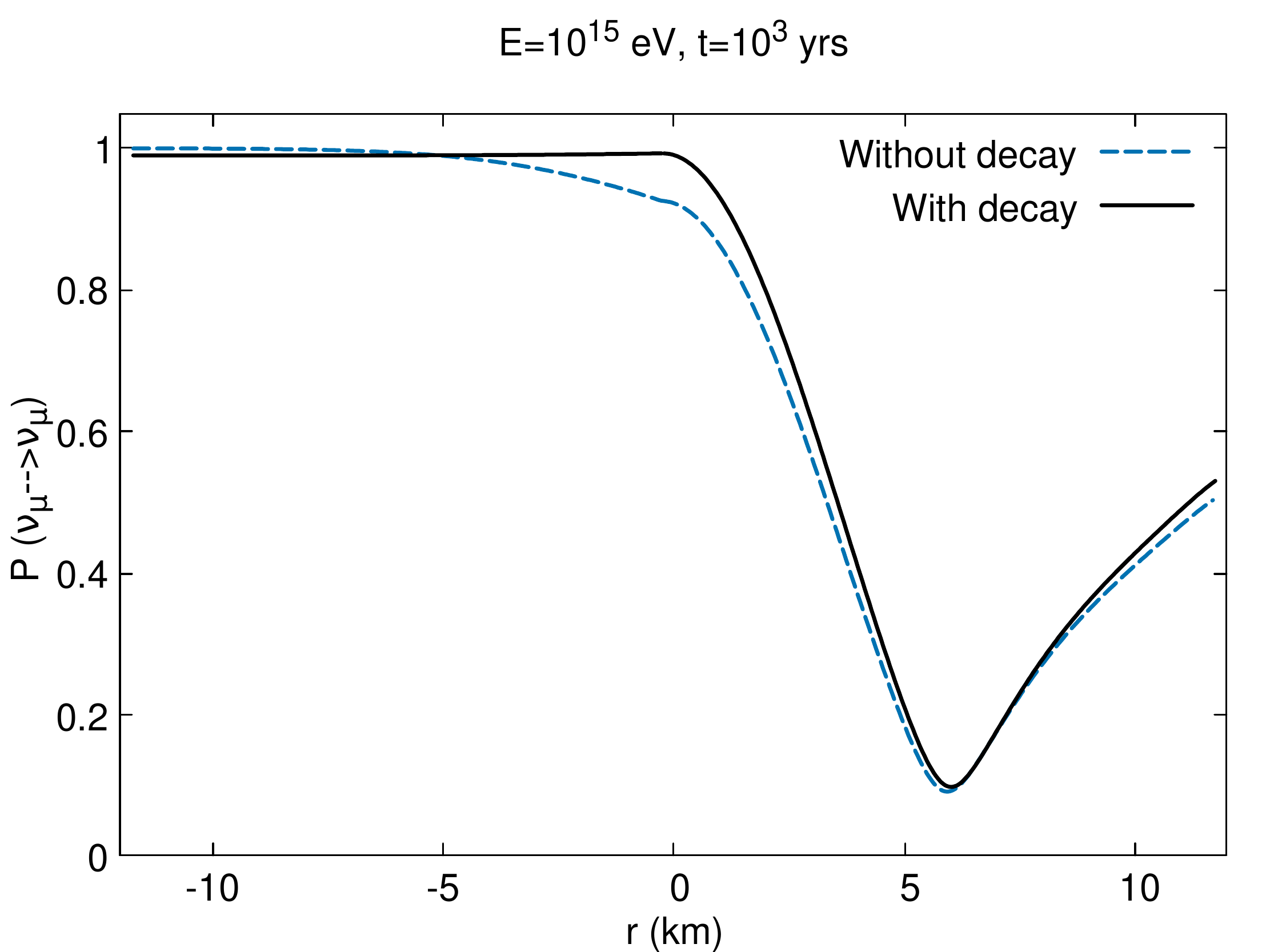}\hspace*{1mm}\includegraphics[scale=0.4]{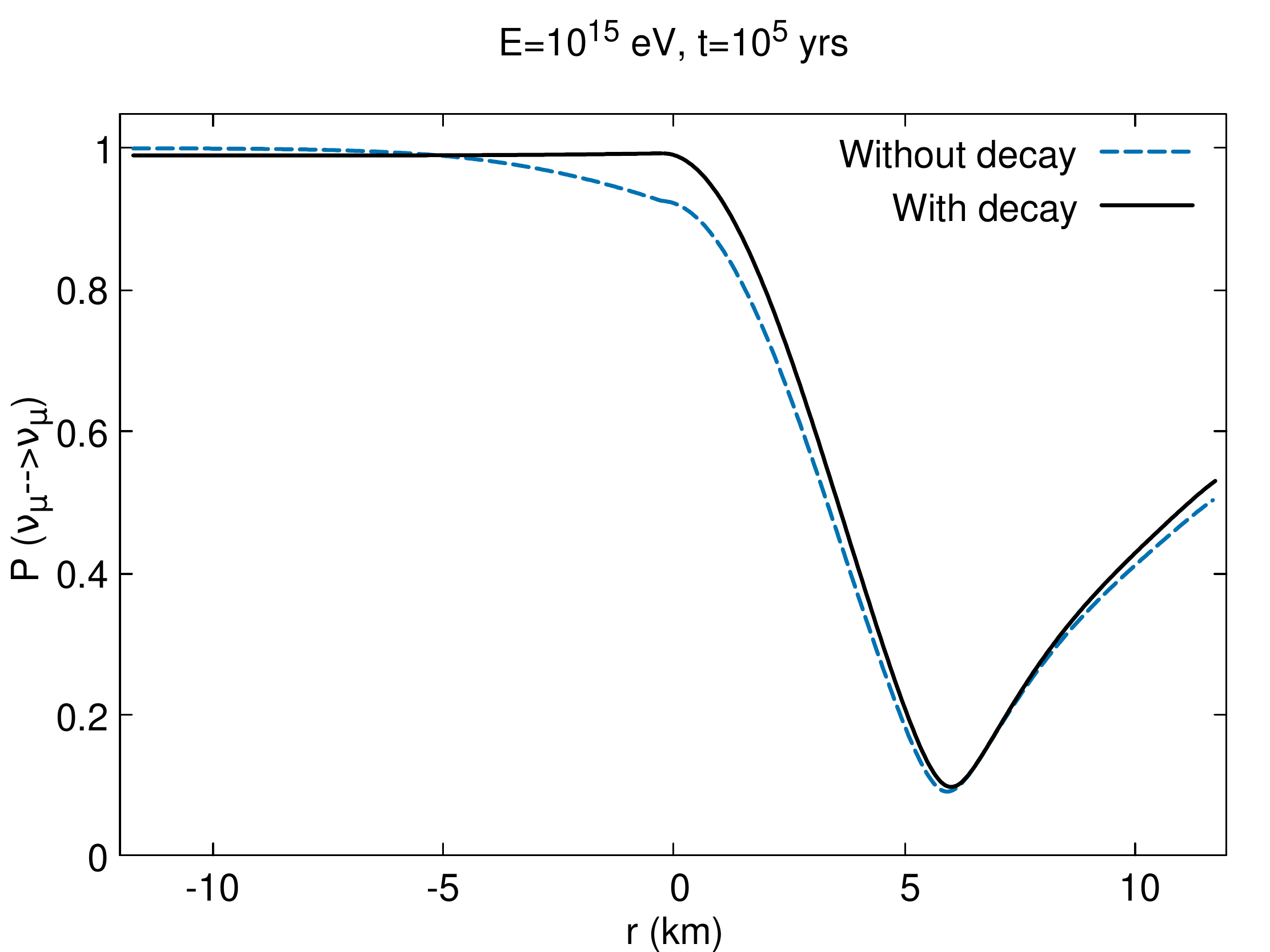}
\caption{Variation of survival probability $P(\nu_{\mu}\rightarrow \nu_{\mu})$ with radial distance for decaying magnetic field  in $npe$ matter of NS.}
\label{f3}
\end{figure}

Fig. \ref{f3} illustrates the muon neutrino survival probability for the case of a decaying magnetic field of a NS for $npe$ matter. For comparison, the plot for a non-decaying field is also shown. We consider a time span of $10^3$ years in which the magnetic field of the magnetar is reduced at least by five orders of magnitude. Further, it should be mentioned that at $t\approx10^3$ years, the core temperature is $\sim 10^8$ K which implies that the ambipolar diffusion could be the most significant process for the dissipation of the magnetic field in the magnetar. It can be observed from the figure that at $t=10^3$ years, around the central region of the star ($R\approx 0$), the decaying magnetic field enhances the survival probability whereas at the point of $\nu_{\mu}$ emission from the NS $i.e.$ at $r=R$, the decaying magnetic field induces slightly higher survival probability ($\sim 0.53$) as compared to that of no decay ($\sim 0.5$). Thus the decaying magnetic field  favours $\sim 5\%$ enhancement in the 
 $\nu_{\mu}$ flux as compared to the temporally static magnetic field. The  results remain almost the same at $t=10^5$ years, as can be seen from the figure. Thus, it can be concluded that there can only be a  marginal alteration in the outgoing neutrino flux  from the NS for a decaying magnetic field as compared to the static field. Further, these conclusions are almost independent of the age of the NS.  The results are further illustrated in table \ref{tt}. 

\begin{center}
\begin{table}
  \begin{tabular}{|c|c|c|}
    \hline
  Quantities  & Without decay & With decay \\
     \hline
$P(\nu_{\mu}\rightarrow \nu_{\mu})$ at $r=R$ & $0.5$ & $0.53$  \\ 
\hline
$\%$ reduction in $\nu_{\mu}$ flux at $r=R$ & $50\%$ & $47\%$ \\
\hline  
     \end{tabular} 
  \caption{Survival probability and the $\%$ reduction of the incoming $\nu_{\mu}$ flux, at the point of emission from the star at $r=R$, $R=11.8$ km.}
  \label{tt}
\end{table}
\end{center}

%%%%%%%%%%%%%%%%%%%%%%%%%%%%%%%%%%
\section{Cosmic neutrinos passing through  white dwarf}
\label{wd}
%%%%%%%%%%%%%%%%%%%%%%%%%%%%%%%%%%

White dwarfs (WD) are another branch of compact stars the formation of which originated from the progenitor stars having lower mass ($1M_{\odot}\leq M \leq 8M_{\odot}$). When there is sufficient nuclear fusion in the core of such a low mass main sequence star which generates helium or some successive heavier nuclei, the outer layer of the star expands to form a red giant (RG). Due to rapid expansion and cooling of the outer layer it becomes unstable and is expelled as a planetary nebulae. The remnant forms WD in which the inward gravitational pull is balanced by the electron degeneracy pressure \cite{2015SSRv..191..111F}. The maximum mass of a WD is about $1.475~M_{\odot}$ which is the Chandrashekhar limit, having its central density $\sim10^6$ gm cm$^{-3}$. The core of a low mass WD usually consists of helium, while heavier ones may contain carbon or oxygen. WDs are smaller compared to NSs and are less massive. Considering non-relativistic degenerate electron gas, the mass radius relation ($M-R$) and the central density are expressed as \cite{Drewes:2021fjx}
\begin{eqnarray}R&=&(10,500~{\rm km})~\left(\frac{0.6M_{\odot}}{M}\right)^{1/3}(2Y_e)^{5/3} \nonumber\\\rho_{central}&=&1.46\times 10^6~{\rm gm~cm^{-3}}\left(\frac{M}{0.6M_{\odot}}\right)^{2}(2Y_e)^{-5},\end{eqnarray} 
where $0.6M_{\odot}$ is the canonical mass of a WD. $Y_e$ is the electron number density per baryon, $Y_e=n_e/(n_p+n_n)$ with $n_e$, $n_p$ and $n_n$ being the number density of electron, proton and neutron, respectively. WDs usually consist of $npe$ matter. Muons cannot be present inside WDs, as the Fermi energy of muons is very small inside the star. Isolated WDs may contain large internal magnetic fields, although the surface value of the magnetic field may be vanishing. In a recent work, the upper bound of the WD magnetic field is estimated to be as \cite{Drewes:2021fjx} \begin{equation}B_{max}=8.8\times 10^{11} G \left(\frac{M}{0.6M_{\odot}}\right)^{1/3}\end{equation}WDs are also capable of producing MeV neutrinos at their core. However, similar to the previous analysis in case of NS, we proceed with UHE Dirac neutrinos coming from a local source which pass through the WD. Due to large magnetic field of WD, it is expected that a fraction of the incoming UHE is converted to their sterile counterpart. 

For our analysis, we consider an isolated WD of mass $0.6~M_{\odot}$ and radius $10,500$ km having constant internal magnetic field of $8.8 \times 10^{11}$ G with $npe$ matter configuration having $Y_e=0.5$. We assume that the local UHE neutrino source gives away $\nu_{\mu}$ beam which crosses the white dwarf along its diameter, entering at the point $r=-10,500$ km considering at its center at $r=0$ and exiting from $r=10,500$ km. We determine the survival probability of $\nu_{\mu}$ after they pass through the star at the point of emission, $r=10,500$ km.

\begin{figure}
\includegraphics[scale=0.4]{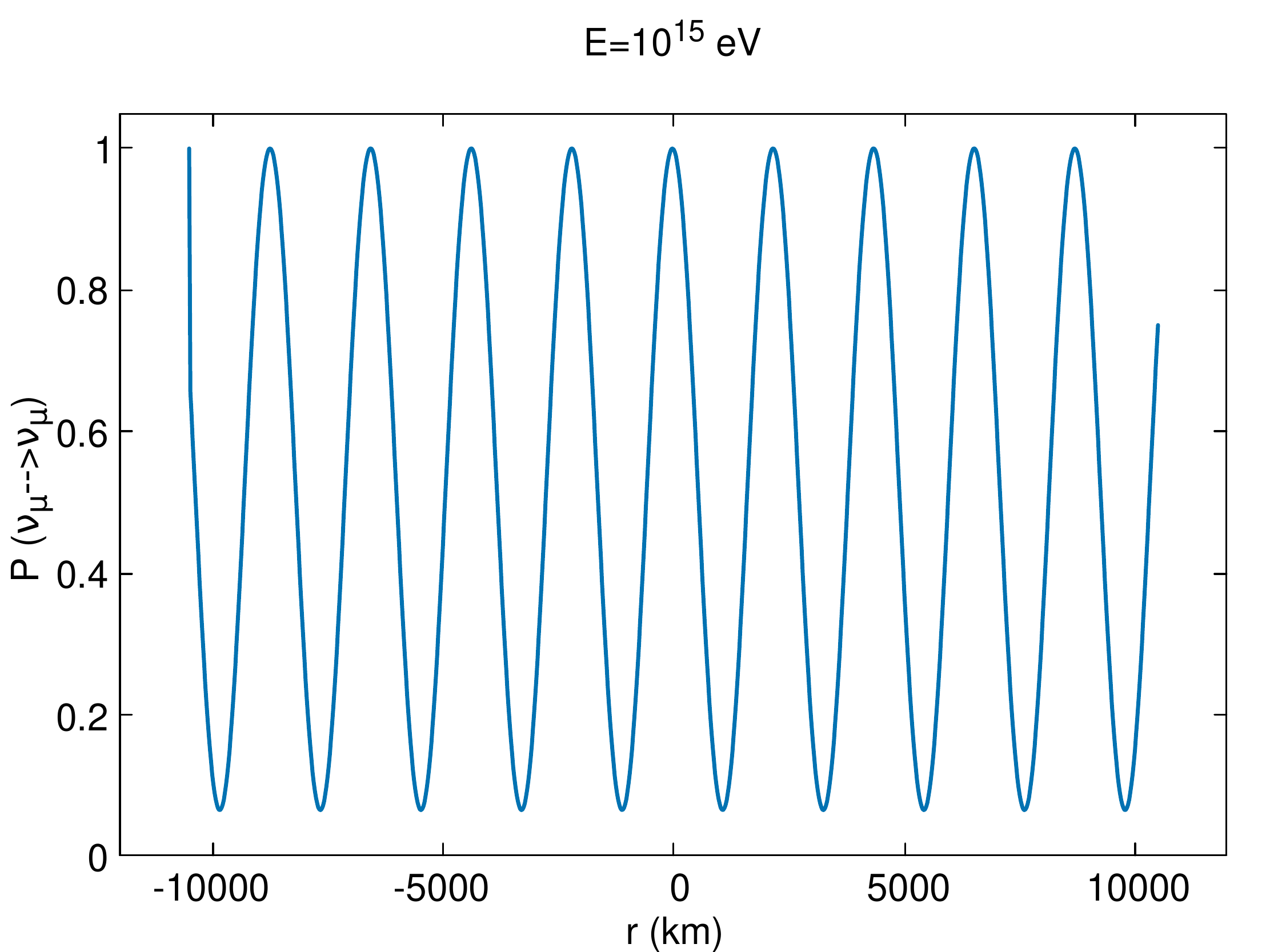}\hspace*{1mm}\includegraphics[scale=0.4]{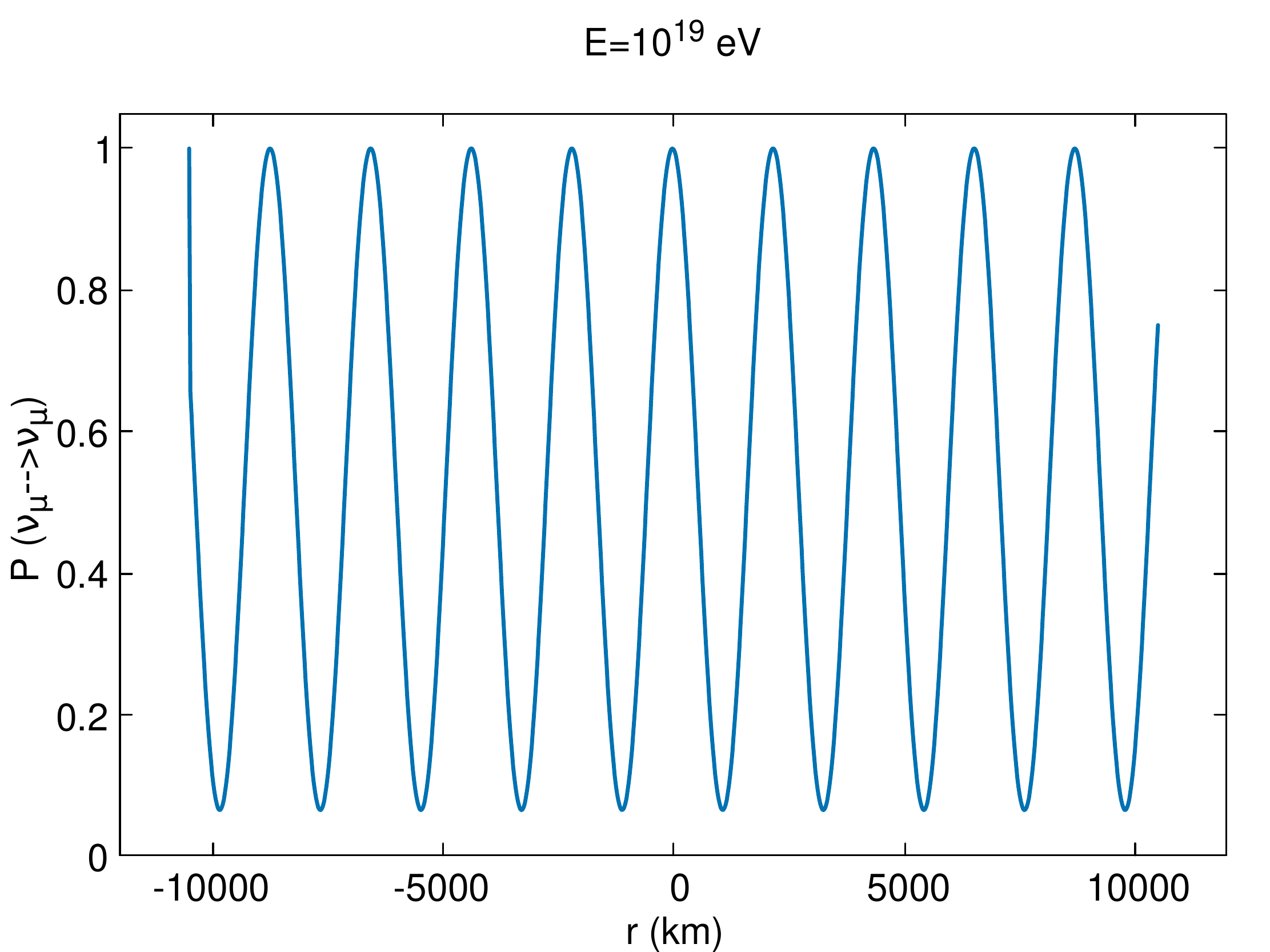}
\caption{Variation of survival probability $P({\nu_{\mu}}\rightarrow {\nu_{\mu}})$ with the radial distance inside the WD star for two different energies.}
\label{f4}
\end{figure}

In fig. \ref{f4}, we represent $\nu_{\mu}$ survival probability varying with the radial distance inside the WD star. Similar to fig. \ref{f1}, in the case of WD, we also present our results considering the energy values of $10^{15}$ eV and $10^{19}$ eV in the left and right panels of the plots, respectively. It is observed that the survival probability follows an oscillatory pattern throughout the star and retains a value of $\sim 0.75$ at the exit point from the star at $r=10,500$ km  implying a  reduction of incoming $\nu_{\mu}$ flux by $25\%$. The result does not vary significantly in case of lower and higher energy values. Compared to NS, UHE Dirac $\nu_{\mu}$s are affected to a lesser extent by the SFP phenomena inside the WD which can be accredited to a smaller value of magnetic field. Thus, in the case of a WD, the conversion of Dirac neutrinos into the sterile form occurs in a lower amount as compared to the NS. This difference can be determined from the proper estimation of the flux by a suitable terrestrial neutrino detector with high efficiency.

Unlike NS, the decay of magnetic fields in a magnetic WD is not realistic as the decay time is very large $\sim  (2-6) \times 10^{11}$ yrs. Therefore, we curtail our analysis to the estimation of flux reduction of UHE neutrinos passing through WD stars containing $npe$ matter and constant magnetic field without delving into the concept of field decay.

%%%%%%%%%%%%%%%%%%%%%%%%%%
\section{Conclusion}\label{con}
%%%%%%%%%%%%%%%%%%%%%%%%%%
We appraise the reduction of flux of ultra high energy Dirac neutrinos passing through two of the most compact stellar objects in the universe, neutron star  and white draft, after being generated from a point source residing close enough to the compact star. As a model star, we consider a neutron star  with mass $2.3~M_{\odot}$ and radius $11.8$ km having a mammoth central magnetic field of $10^{18}$ G. Using the current experimental bound on the magnetic moment of neutrinos and assuming GM1 parametrization, we evaluate the survival probability of the incoming UHE  $\nu_{\mu}$ beam at the point of emission for $npe$ as well as $npe\mu$ matter. We observe the following:

\begin{itemize}

\item For a non-decaying magnetic field, the flux of a $\nu_{\mu}$ beam can be reduced to half at the emission point for $npe$ matter.

\item For  muons in the neutron star matter, the survival probability of $\nu_{\mu}$ decreases from 0.50 to 0.45.

\item These results remain unaltered in the  $\nu_{\mu}$ energy range of ($10^{15}- 10^{19}$) eV.

\item For a decaying magnetic field in  $npe$ matter of neutron star, the survival probability of $\nu_{\mu}$ increases by $\sim 5\%$ as compared to the  temporally static field. Although the decaying and non-decaying scenarios show a significant difference in the values of neutrino survival probability around the central region of the neutron star, this may not have any observational implications.

\end{itemize}

We then consider a  white dwarf star consisting of only $npe$ matter and having mass $0.6M_{\odot}$ and radius $10,500$ km with a constant magnetic field of $8.8\times 10^{11}$ G. This leads to $\nu_{\mu}$ survival probability of $\sim 0.8$ at the point of emission after they traverse along its diameter. Like neutron stars, the conclusions are almost the same in the energy range of  ($10^{15}- 10^{19}$) eV.  We do not consider the case of magnetic field evolution in white dwarf due to its unusually large decaying time scale.


\begin{thebibliography}{99}

\bibitem{IceCube:2018cha}
M.~G.~Aartsen \textit{et al.} [IceCube],
``Neutrino emission from the direction of the blazar TXS 0506+056 prior to the IceCube-170922A alert,''
Science \textbf{361} (2018) no.6398, 147-151
%doi:10.1126/science.aat2890
[arXiv:1807.08794 [astro-ph.HE]].

\bibitem{Ackermann:2022rqc}
M.~Ackermann, S.~K.~Agarwalla, J.~Alvarez-Mu\~niz, R.~Alves Batista, C.~A.~Arg\"uelles, M.~Bustamante, B.~A.~Clark, A.~Cummings, S.~Das and V.~Decoene, \textit{et al.}
``High-Energy and Ultra-High-Energy Neutrinos,''
[arXiv:2203.08096 [hep-ph]].

\bibitem{Hussain:2006wg}
S.~Hussain, D.~Marfatia, D.~W.~McKay and D.~Seckel,
``Cross section dependence of event rates at neutrino telescopes,''
Phys. Rev. Lett. \textbf{97} (2006), 161101
%doi:10.1103/PhysRevLett.97.161101
[arXiv:hep-ph/0606246 [hep-ph]].

\bibitem{Kusenko:2001gj}
A.~Kusenko and T.~J.~Weiler,
``Neutrino cross-sections at high-energies and the future observations of ultrahigh-energy cosmic rays,''
Phys. Rev. Lett. \textbf{88} (2002), 161101
%doi:10.1103/PhysRevLett.88.161101
[arXiv:hep-ph/0106071 [hep-ph]].

\bibitem{Hooper:2002yq}
D.~Hooper,
``Measuring high-energy neutrino nucleon cross-sections with future neutrino telescopes,''
Phys. Rev. D \textbf{65} (2002), 097303
%doi:10.1103/PhysRevD.65.097303
[arXiv:hep-ph/0203239 [hep-ph]].

\bibitem{Friess:2002cc}
J.~J.~Friess, T.~Han and D.~Hooper,
``TeV string state excitation via high-energy cosmic neutrinos,''
Phys. Lett. B \textbf{547} (2002), 31-36
%doi:10.1016/S0370-2693(02)02728-4
[arXiv:hep-ph/0204112 [hep-ph]].

\bibitem{Anchordoqui:2002vb}
L.~A.~Anchordoqui, J.~L.~Feng, H.~Goldberg and A.~D.~Shapere,
``Neutrino bounds on astrophysical sources and new physics,''
Phys. Rev. D \textbf{66} (2002), 103002
%doi:10.1103/PhysRevD.66.103002
[arXiv:hep-ph/0207139 [hep-ph]].

\bibitem{Anchordoqui:2005pn}
L.~A.~Anchordoqui, J.~L.~Feng and H.~Goldberg,
``Particle physics on ice: Constraints on neutrino interactions far above the weak scale,''
Phys. Rev. Lett. \textbf{96} (2006), 021101
%doi:10.1103/PhysRevLett.96.021101
[arXiv:hep-ph/0504228 [hep-ph]].

\bibitem{Bustamante:2017xuy}
M.~Bustamante and A.~Connolly,
``Extracting the Energy-Dependent Neutrino-Nucleon Cross Section above 10 TeV Using IceCube Showers,''
Phys. Rev. Lett. \textbf{122} (2019) no.4, 041101
%doi:10.1103/PhysRevLett.122.041101
[arXiv:1711.11043 [astro-ph.HE]].


\bibitem{IceCube:2017roe}
M.~G.~Aartsen \textit{et al.} [IceCube],
``Measurement of the multi-TeV neutrino cross section with IceCube using Earth absorption,''
Nature \textbf{551} (2017), 596-600
%doi:10.1038/nature24459
[arXiv:1711.08119 [hep-ex]]


\bibitem{IceCube:2020rnc}
R.~Abbasi \textit{et al.} [IceCube],
``Measurement of the high-energy all-flavor neutrino-nucleon cross section with IceCube,''
%doi:10.1103/PhysRevD.104.022001
[arXiv:2011.03560 [hep-ex]].


\bibitem{Halzen:2010yj}
F.~Halzen and S.~R.~Klein,
``IceCube: An Instrument for Neutrino Astronomy,''
Rev. Sci. Instrum. \textbf{81} (2010), 081101
%doi:10.1063/1.3480478
[arXiv:1007.1247 [astro-ph.HE]].


\bibitem{Gaisser:2014foa}
T.~Gaisser and F.~Halzen,
``IceCube,''
Ann. Rev. Nucl. Part. Sci. \textbf{64} (2014), 101-123
%doi:10.1146/annurev-nucl-102313-025321

\bibitem{KM3Net:2016zxf}
S.~Adrian-Martinez \textit{et al.} [KM3Net],
``Letter of intent for KM3NeT 2.0,''
J. Phys. G \textbf{43} (2016) no.8, 084001
%doi:10.1088/0954-3899/43/8/084001
[arXiv:1601.07459 [astro-ph.IM]]

\bibitem{Baikal-GVD:2019kwy}
A.~D.~Avrorin \textit{et al.} [Baikal-GVD],
``Neutrino Telescope in Lake Baikal: Present and Future,''
PoS \textbf{ICRC2019} (2020), 1011
%doi:10.22323/1.358.1011
[arXiv:1908.05427 [astro-ph.HE]].

\bibitem{P-ONE:2020ljt}
M.~Agostini \textit{et al.} [P-ONE],
``The Pacific Ocean Neutrino Experiment,''
Nature Astron. \textbf{4} (2020) no.10, 913-915
%doi:10.1038/s41550-020-1182-4
[arXiv:2005.09493 [astro-ph.HE]].

\bibitem{IceCube-Gen2:2020qha}
M.~G.~Aartsen \textit{et al.} [IceCube-Gen2],
``IceCube-Gen2: the window to the extreme Universe,''
J. Phys. G \textbf{48} (2021) no.6, 060501
%doi:10.1088/1361-6471/abbd48
[arXiv:2008.04323 [astro-ph.HE]].

\bibitem{Prohira:2019glh}
S.~Prohira, K.~D.~de Vries, P.~Allison, J.~Beatty, D.~Besson, A.~Connolly, N.~van Eijndhoven, C.~Hast, C.~Y.~Kuo and U.~A.~Latif, \textit{et al.}
``Observation of Radar Echoes From High-Energy Particle Cascades,''
Phys. Rev. Lett. \textbf{124} (2020) no.9, 091101
%doi:10.1103/PhysRevLett.124.091101
[arXiv:1910.12830 [astro-ph.HE]].

\bibitem{Romero-Wolf:2020pzh}
A.~Romero-Wolf, J.~Alvarez-Mu\~niz, W.~R.~Carvalho, A.~Cummings, H.~Schoorlemmer, S.~Wissel, E.~Zas, C.~Arg\"uelles, H.~Barreda and J.~Bazo, \textit{et al.}
``An Andean Deep-Valley Detector for High-Energy Tau Neutrinos,''
[arXiv:2002.06475 [astro-ph.IM]].

\bibitem{Brown:2021lef}
A.~M.~Brown, M.~Bagheri, M.~Doro, E.~Gazda, D.~Kieda, C.~Lin, Y.~Onel, N.~Otte, I.~Taboada and A.~Wang,
``Trinity: An Imaging Air Cherenkov Telescope to Search for Ultra-High-Energy Neutrinos,''
[arXiv:2109.03125 [astro-ph.IM]].

\bibitem{Brown:2021ane}
A.~M.~Brown, M.~Bagheri, M.~Doro, E.~Gazda, D.~Kieda, C.~Lin, N.~Otte, I.~Taboada and A.~Wang,
``Trinity: an imaging air Cherenkov telescope to search for Ultra-High-Energy neutrinos,''
PoS \textbf{ICRC2021} (2021), 1179
%doi:10.22323/1.395.1179


\bibitem{Connolly:2011vc}
A.~Connolly, R.~S.~Thorne and D.~Waters,
``Calculation of High Energy Neutrino-Nucleon Cross Sections and Uncertainties Using the MSTW Parton Distribution Functions and Implications for Future Experiments,''
Phys. Rev. D \textbf{83} (2011), 113009
%doi:10.1103/PhysRevD.83.113009
[arXiv:1102.0691 [hep-ph]].

\bibitem{Garcia:2020jwr}
A.~Garcia, R.~Gauld, A.~Heijboer and J.~Rojo,
``Complete predictions for high-energy neutrino propagation in matter,''
JCAP \textbf{09} (2020), 025
%doi:10.1088/1475-7516/2020/09/025
[arXiv:2004.04756 [hep-ph]].

\bibitem{Gandhi:1995tf}
R.~Gandhi, C.~Quigg, M.~H.~Reno and I.~Sarcevic,
``Ultrahigh-energy neutrino interactions,''
Astropart. Phys. \textbf{5} (1996), 81-110
%doi:10.1016/0927-6505(96)00008-4
[arXiv:hep-ph/9512364 [hep-ph]].

\bibitem{Gandhi:1998ri}
R.~Gandhi, C.~Quigg, M.~H.~Reno and I.~Sarcevic,
``Neutrino interactions at ultrahigh-energies,''
Phys. Rev. D \textbf{58} (1998), 093009
%doi:10.1103/PhysRevD.58.093009
[arXiv:hep-ph/9807264 [hep-ph]].

\bibitem{Palomares-Ruiz:2005npx}
S.~Palomares-Ruiz, A.~Irimia and T.~J.~Weiler,
``Acceptances for space-based and ground-based fluorescence detectors, and inference of the neutrino-nucleon cross-section above 10**19-ev,''
Phys. Rev. D \textbf{73} (2006), 083003
%doi:10.1103/PhysRevD.73.083003
[arXiv:astro-ph/0512231 [astro-ph]].

\bibitem{Venters:2019xwi}
T.~M.~Venters, M.~H.~Reno, J.~F.~Krizmanic, L.~A.~Anchordoqui, C.~Gu\'epin and A.~V.~Olinto,
``POEMMA's Target of Opportunity Sensitivity to Cosmic Neutrino Transient Sources,''
Phys. Rev. D \textbf{102} (2020), 123013
%doi:10.1103/PhysRevD.102.123013
[arXiv:1906.07209 [astro-ph.HE]].

\bibitem{Reno:2019jtr}
M.~H.~Reno, J.~F.~Krizmanic and T.~M.~Venters,
``Cosmic tau neutrino detection via Cherenkov signals from air showers from Earth-emerging taus,''
Phys. Rev. D \textbf{100} (2019) no.6, 063010
%doi:10.1103/PhysRevD.100.063010
[arXiv:1902.11287 [astro-ph.HE]].



\bibitem{Aramo:2004pr}
C.~Aramo, A.~Insolia, A.~Leonardi, G.~Miele, L.~Perrone, O.~Pisanti and D.~V.~Semikoz,
``Earth-skimming UHE Tau neutrinos at the fluorescence detector of Pierre Auger observatory,''
Astropart. Phys. \textbf{23} (2005), 65-77
%doi:10.1016/j.astropartphys.2004.11.008
[arXiv:astro-ph/0407638 [astro-ph]]

\bibitem{Nitz:2021kdx}
D.~F.~Nitz [Pierre Auger],
``New Electronics for the Surface Detectors of the Pierre Auger Observatory,''
PoS \textbf{ICRC2019} (2021), 370
%doi:10.22323/1.358.0370

\bibitem{PierreAuger:2021ccl}
G.~Cataldi \textit{et al.} [Pierre Auger],
``The upgrade of the Pierre Auger Observatory with the Scintillator Surface Detector,''
PoS \textbf{ICRC2021} (2021), 251
%doi:10.22323/1.395.0251

\bibitem{ANITA:2019wyx}
P.~W.~Gorham \textit{et al.} [ANITA],
``Constraints on the ultrahigh-energy cosmic neutrino flux from the fourth flight of ANITA,''
Phys. Rev. D \textbf{99} (2019) no.12, 122001
%doi:10.1103/PhysRevD.99.122001
[arXiv:1902.04005 [astro-ph.HE]].

\bibitem{ARIANNA:2019scz}
A.~Anker \textit{et al.} [ARIANNA],
``Targeting ultra-high energy neutrinos with the ARIANNA experiment,''
Adv. Space Res. \textbf{64} (2019), 2595-2609
%doi:10.1016/j.asr.2019.06.016
[arXiv:1903.01609 [astro-ph.IM]].

\bibitem{ARA:2019wcf}
P.~Allison \textit{et al.} [ARA],
``Constraints on the diffuse flux of ultrahigh energy neutrinos from four years of Askaryan Radio Array data in two stations,''
Phys. Rev. D \textbf{102} (2020) no.4, 043021
%doi:10.1103/PhysRevD.102.043021
[arXiv:1912.00987 [astro-ph.HE]].

\bibitem{ARA:2022rwq}
P.~Allison \textit{et al.} [ARA],
``Low-threshold ultrahigh-energy neutrino search with the Askaryan Radio Array,''
Phys. Rev. D \textbf{105} (2022) no.12, 122006
%doi:10.1103/PhysRevD.105.122006
[arXiv:2202.07080 [astro-ph.HE]].


\bibitem{RNO-G:2020rmc}
J.~A.~Aguilar \textit{et al.} [RNO-G],
``Design and Sensitivity of the Radio Neutrino Observatory in Greenland (RNO-G),''
JINST \textbf{16} (2021) no.03, P03025
%doi:10.1088/1748-0221/16/03/P03025
[arXiv:2010.12279 [astro-ph.IM]].

\bibitem{RNO-G:2021hfx}
J.~A.~Aguilar \textit{et al.} [RNO-G],
``The Radio Neutrino Observatory Greenland (RNO-G),''
PoS \textbf{ICRC2021} (2021), 001
%doi:10.22323/1.395.0001

\bibitem{PUEO:2020bnn}
Q.~Abarr \textit{et al.} [PUEO],
``The Payload for Ultrahigh Energy Observations (PUEO): a white paper,''
JINST \textbf{16} (2021) no.08, 08
%doi:10.1088/1748-0221/16/08/P08035
[arXiv:2010.02892 [astro-ph.IM]].

\bibitem{Giunti:2014ixa}
C.~Giunti and A.~Studenikin,
``Neutrino electromagnetic interactions: a window to new physics,''
Rev. Mod. Phys. \textbf{87}, 531 (2015)
%doi:10.1103/RevModPhys.87.531
[arXiv:1403.6344 [hep-ph]].

%\cite{Broggini:2012df}
\bibitem{Broggini:2012df}
C.~Broggini, C.~Giunti and A.~Studenikin,
``Electromagnetic Properties of Neutrinos,''
Adv. High Energy Phys. \textbf{2012}, 459526 (2012)
%doi:10.1155/2012/459526
[arXiv:1207.3980 [hep-ph]].
%113 citations counted in INSPIRE as of 03 Aug 2022

\bibitem{Alok:2022pdn}
A.~K.~Alok, N.~R.~S.~Chundawat and A.~Mandal,
``Cosmic neutrino flux and spin flavor oscillations in intergalactic medium,''
[arXiv:2207.13034 [hep-ph]].


\bibitem{Bahcall:2004ut}
J.~N.~Bahcall, M.~C.~Gonzalez-Garcia and C.~Pena-Garay,
``Solar neutrinos before and after neutrino 2004,''
JHEP \textbf{08} (2004), 016
%doi:10.1088/1126-6708/2004/08/016
[arXiv:hep-ph/0406294 [hep-ph]].

\bibitem{KamLAND:2002uet}
K.~Eguchi \textit{et al.} [KamLAND],
``First results from KamLAND: Evidence for reactor anti-neutrino disappearance,''
Phys. Rev. Lett. \textbf{90} (2003), 021802
%doi:10.1103/PhysRevLett.90.021802
[arXiv:hep-ex/0212021 [hep-ex]].

\bibitem{KamLAND:2004mhv}
T.~Araki \textit{et al.} [KamLAND],
``Measurement of neutrino oscillation with KamLAND: Evidence of spectral distortion,''
Phys. Rev. Lett. \textbf{94} (2005), 081801
%doi:10.1103/PhysRevLett.94.081801
[arXiv:hep-ex/0406035 [hep-ex]]


\bibitem{Super-Kamiokande:2004orf}
Y.~Ashie \textit{et al.} [Super-Kamiokande],
``Evidence for an oscillatory signature in atmospheric neutrino oscillation,''
Phys. Rev. Lett. \textbf{93} (2004), 101801
%doi:10.1103/PhysRevLett.93.101801
[arXiv:hep-ex/0404034 [hep-ex]].

\bibitem{MINOS:2006foh}
D.~G.~Michael \textit{et al.} [MINOS],
``Observation of muon neutrino disappearance with the MINOS detectors and the NuMI neutrino beam,''
Phys. Rev. Lett. \textbf{97} (2006), 191801
%doi:10.1103/PhysRevLett.97.191801
[arXiv:hep-ex/0607088 [hep-ex]].

\bibitem{T2K:2013ppw}
K.~Abe \textit{et al.} [T2K],
``Observation of Electron Neutrino Appearance in a Muon Neutrino Beam,''
Phys. Rev. Lett. \textbf{112} (2014), 061802
%doi:10.1103/PhysRevLett.112.061802
[arXiv:1311.4750 [hep-ex]].




\bibitem{T2K:2013bzi}
K.~Abe \textit{et al.} [T2K],
``Measurement of Neutrino Oscillation Parameters from Muon Neutrino Disappearance with an Off-axis Beam,''
Phys. Rev. Lett. \textbf{111} (2013) no.21, 211803
%doi:10.1103/PhysRevLett.111.211803
[arXiv:1308.0465 [hep-ex]].


%\cite{Fujikawa:1980yx}
\bibitem{Fujikawa:1980yx}
K.~Fujikawa and R.~Shrock,
``The Magnetic Moment of a Massive Neutrino and Neutrino Spin Rotation,''
Phys. Rev. Lett. \textbf{45}, 963 (1980)
%doi:10.1103/PhysRevLett.45.963
%507 citations counted in INSPIRE as of 03 Aug 2022

\bibitem{KATRIN:2021uub}
M.~Aker \textit{et al.} [KATRIN],
``Direct neutrino-mass measurement with sub-electronvolt sensitivity,''
Nature Phys. \textbf{18} (2022) no.2, 160-166
%doi:10.1038/s41567-021-01463-1
[arXiv:2105.08533 [hep-ex]].

\bibitem{Povarov:2007zz}
A.~V.~Povarov,
``Scalar-leptoquark contributions to the neutrino magnetic moment,''
Phys. Atom. Nucl. \textbf{70}, 871-878 (2007)
%doi:10.1134/S1063778807050109

\bibitem{Sanchez-Velez:2022nwm}
R.~S\'anchez-V\'elez,
``Neutrino electromagnetic properties in a Leptoquark model,''
[arXiv:2204.01167 [hep-ph]].

\bibitem{Chua:1998yk}
C.~K.~Chua and W.~Y.~P.~Hwang,
``The Neutrino magnetic moment induced by leptoquarks,''
Phys. Rev. D \textbf{60} (1999), 073002
%doi:10.1103/PhysRevD.60.073002
[arXiv:hep-ph/9811232 [hep-ph]].

%\cite{Voloshin:1986ty}
\bibitem{Voloshin:1986ty}
M.~B.~Voloshin and M.~I.~Vysotsky,
``Neutrino Magnetic Moment and Time Variation of Solar Neutrino Flux,''
Sov. J. Nucl. Phys. \textbf{44}, 544 (1986)
ITEP-1-1986.
%218 citations counted in INSPIRE as of 03 Aug 2022


%\cite{Aboubrahim:2013yfa}
\bibitem{Aboubrahim:2013yfa}
A.~Aboubrahim, T.~Ibrahim, A.~Itani and P.~Nath,
``Large Neutrino Magnetic Dipole Moments in MSSM Extensions,''
Phys. Rev. D \textbf{89}, no.5, 055009 (2014)
%doi:10.1103/PhysRevD.89.055009
[arXiv:1312.2505 [hep-ph]].
%21 citations counted in INSPIRE as of 03 Aug 2022

%\cite{Fukuyama:2003uz}
\bibitem{Fukuyama:2003uz}
T.~Fukuyama, T.~Kikuchi and N.~Okada,
``Neutrino magnetic moments and minimal supersymmetric SO(10) model,''
Int. J. Mod. Phys. A \textbf{19}, 4825-4834 (2004)
%doi:10.1142/S0217751X04020038
[arXiv:hep-ph/0306025 [hep-ph]].
%9 citations counted in INSPIRE as of 03 Aug 2022


%\cite{Babu:2020ivd}
\bibitem{Babu:2020ivd}
K.~S.~Babu, S.~Jana and M.~Lindner,
``Large Neutrino Magnetic Moments in the Light of Recent Experiments,''
JHEP \textbf{10}, 040 (2020)
%doi:10.1007/JHEP10(2020)040
[arXiv:2007.04291 [hep-ph]].
%54 citations counted in INSPIRE as of 03 Aug 2022



%\cite{Healey:2013vka}
\bibitem{Healey:2013vka}
K.~J.~Healey, A.~A.~Petrov and D.~Zhuridov,
``Nonstandard neutrino interactions and transition magnetic moments,''
Phys. Rev. D \textbf{87}, no.11, 117301 (2013)
[erratum: Phys. Rev. D \textbf{89}, no.5, 059904 (2014)]
%doi:10.1103/PhysRevD.87.117301
[arXiv:1305.0584 [hep-ph]].
%20 citations counted in INSPIRE as of 03 Aug 2022


%\cite{Papoulias:2015iga}
\bibitem{Papoulias:2015iga}
D.~K.~Papoulias and T.~S.~Kosmas,
``Neutrino transition magnetic moments within the non-standard neutrino\textendash{}nucleus interactions,''
Phys. Lett. B \textbf{747}, 454-459 (2015)
%doi:10.1016/j.physletb.2015.06.039
[arXiv:1506.05406 [hep-ph]].
%20 citations counted in INSPIRE as of 03 Aug 2022

%\cite{Kharlanov:2020cti}
\bibitem{Kharlanov:2020cti}
O.~G.~Kharlanov and P.~I.~Shustov,
``Effects of nonstandard neutrino self-interactions and magnetic moment on collective Majorana neutrino oscillations,''
Phys. Rev. D \textbf{103}, no.9, 095004 (2021)
%doi:10.1103/PhysRevD.103.095004
[arXiv:2010.05329 [hep-ph]].
%4 citations counted in INSPIRE as of 03 Aug 2022

%\cite{Beda:2009kx}
\bibitem{Beda:2009kx}
A.~G.~Beda, E.~V.~Demidova, A.~S.~Starostin, V.~B.~Brudanin, V.~G.~Egorov, D.~V.~Medvedev, M.~V.~Shirchenko and T.~Vylov,
``GEMMA experiment: Three years of the search for the neutrino magnetic moment,''
Phys. Part. Nucl. Lett. \textbf{7}, 406-409 (2010)
%doi:10.1134/S1547477110060063
[arXiv:0906.1926 [hep-ex]].
%65 citations counted in INSPIRE as of 03 Aug 2022

%\cite{TEXONO:2006xds}
\bibitem{TEXONO:2006xds}
H.~T.~Wong \textit{et al.} [TEXONO],
``A Search of Neutrino Magnetic Moments with a High-Purity Germanium Detector at the Kuo-Sheng Nuclear Power Station,''
Phys. Rev. D \textbf{75}, 012001 (2007)
%doi:10.1103/PhysRevD.75.012001
[arXiv:hep-ex/0605006 [hep-ex]].
%216 citations counted in INSPIRE as of 03 Aug 2022

%\cite{Derbin:1993wy}
\bibitem{Derbin:1993wy}
A.~I.~Derbin, A.~V.~Chernyi, L.~A.~Popeko, V.~N.~Muratova, G.~A.~Shishkina and S.~I.~Bakhlanov,
``Experiment on anti-neutrino scattering by electrons at a reactor of the Rovno nuclear power plant,''
JETP Lett. \textbf{57}, 768-772 (1993)
%107 citations counted in INSPIRE as of 03 Aug 2022

%\cite{Allen:1992qe}
\bibitem{Allen:1992qe}
R.~C.~Allen, H.~H.~Chen, P.~J.~Doe, R.~Hausammann, W.~P.~Lee, X.~Q.~Lu, H.~J.~Mahler, M.~E.~Potter, K.~C.~Wang and T.~J.~Bowles, \textit{et al.}
``Study of electron-neutrino electron elastic scattering at LAMPF,''
Phys. Rev. D \textbf{47}, 11-28 (1993)
%doi:10.1103/PhysRevD.47.11
%155 citations counted in INSPIRE as of 03 Aug 2022

%\cite{LSND:2001akn}
\bibitem{LSND:2001akn}
L.~B.~Auerbach \textit{et al.} [LSND],
``Measurement of electron - neutrino - electron elastic scattering,''
Phys. Rev. D \textbf{63}, 112001 (2001)
%doi:10.1103/PhysRevD.63.112001
[arXiv:hep-ex/0101039 [hep-ex]].
%287 citations counted in INSPIRE as of 03 Aug 2022

%\cite{Heger:2008er}
\bibitem{Heger:2008er}
A.~Heger, A.~Friedland, M.~Giannotti and V.~Cirigliano,
``The Impact of Neutrino Magnetic Moments on the Evolution of Massive Stars,''
Astrophys. J. \textbf{696}, 608-619 (2009)
%doi:10.1088/0004-637X/696/1/608
[arXiv:0809.4703 [astro-ph]].
%45 citations counted in INSPIRE as of 03 Aug 2022

%\cite{Borisov:2014cqa}
\bibitem{Borisov:2014cqa}
A.~V.~Borisov and P.~E.~Sizin,
``Plasmon decay to a neutrino pair via neutrino electromagnetic moments in a strongly magnetized medium,''
%doi:10.1142/9789814663618\_0024
[arXiv:1406.3301 [hep-ph]].
%0 citations counted in INSPIRE as of 03 Aug 2022

%\cite{deGouvea:2012hg}
\bibitem{deGouvea:2012hg}
A.~de Gouvea and S.~Shalgar,
``Effect of Transition Magnetic Moments on Collective Supernova Neutrino Oscillations,''
JCAP \textbf{10}, 027 (2012)
%doi:10.1088/1475-7516/2012/10/027
[arXiv:1207.0516 [astro-ph.HE]].
%70 citations counted in INSPIRE as of 03 Aug 2022

%\cite{Vassh:2015yza}
\bibitem{Vassh:2015yza}
N.~Vassh, E.~Grohs, A.~B.~Balantekin and G.~M.~Fuller,
``Majorana Neutrino Magnetic Moment and Neutrino Decoupling in Big Bang Nucleosynthesis,''
Phys. Rev. D \textbf{92}, no.12, 125020 (2015)
%doi:10.1103/PhysRevD.92.125020
[arXiv:1510.00428 [astro-ph.CO]].
%19 citations counted in INSPIRE as of 03 Aug 2022


%\cite{Viaux:2013hca}
\bibitem{Viaux:2013hca}
N.~Viaux, M.~Catelan, P.~B.~Stetson, G.~Raffelt, J.~Redondo, A.~A.~R.~Valcarce and A.~Weiss,
``Particle-physics constraints from the globular cluster M5: Neutrino Dipole Moments,''
Astron. Astrophys. \textbf{558}, A12 (2013)
%doi:10.1051/0004-6361/201322004
[arXiv:1308.4627 [astro-ph.SR]].
%67 citations counted in INSPIRE as of 03 Aug 2022


%\cite{Joshi:2019dcj}
\bibitem{Joshi:2019dcj}
S.~Joshi and S.~R.~Jain,
``Neutrino spin-flavor oscillations in solar environment,''
%doi:10.1088/1674-4527/20/8/123
[arXiv:1906.09475 [hep-ph]].
%1 citations counted in INSPIRE as of 03 Aug 2022


%\cite{Lattimer:2004pg}
\bibitem{Lattimer:2004pg}
J.~M.~Lattimer and M.~Prakash,
``The physics of neutron stars,''
Science \textbf{304}, 536-542 (2004)
%doi:10.1126/science.1090720
[arXiv:astro-ph/0405262 [astro-ph]].
%952 citations counted in INSPIRE as of 12 Aug 2022

%\cite{Nattila:2017wtj}
\bibitem{Nattila:2017wtj}
J.~N\"attil\"a, M.~C.~Miller, A.~W.~Steiner, J.~J.~E.~Kajava, V.~F.~Suleimanov and J.~Poutanen,
``Neutron star mass and radius measurements from atmospheric model fits to X-ray burst cooling tail spectra,''
Astron. Astrophys. \textbf{608}, A31 (2017)
%doi:10.1051/0004-6361/201731082
[arXiv:1709.09120 [astro-ph.HE]].
%109 citations counted in INSPIRE as of 12 Aug 2022

%\cite{Zhang:2020wov}
\bibitem{Zhang:2020wov}
N.~B.~Zhang and B.~A.~Li,
``Constraints on the muon fraction and density profile in neutron stars,''
Astrophys. J. \textbf{893}, 61 (2020)
%doi:10.3847/1538-4357/ab7dbc
[arXiv:2002.06446 [astro-ph.HE]].
%12 citations counted in INSPIRE as of 12 Aug 2022

%\cite{Esposito:2018gvp}
\bibitem{Esposito:2018gvp}
P.~Esposito, N.~Rea and G.~L.~Israel,
%``Magnetars: a short review and some sparse considerations,''
Astrophys. Space Sci. Libr. \textbf{461}, 97-142 (2020)
%doi:10.1007/978-3-662-62110-3\_3
[arXiv:1803.05716 [astro-ph.HE]].
%35 citations counted in INSPIRE as of 12 Aug 2022

%\cite{Meszaros:2017fcs}
\bibitem{Meszaros:2017fcs}
P.~Meszaros,
``Astrophysical Sources of High Energy Neutrinos in the IceCube Era,''
Ann. Rev. Nucl. Part. Sci. \textbf{67}, 45-67 (2017)
doi:10.1146/annurev-nucl-101916-123304
[arXiv:1708.03577 [astro-ph.HE]].
%58 citations counted in INSPIRE as of 12 Aug 2022

%\cite{Adhikary:2022phm}
\bibitem{Adhikary:2022phm}
J.~Adhikary, A.~K.~Alok, A.~Mandal, T.~Sarkar and S.~Sharma,
``Neutrino spin-flavour precession in magnetized white dwarf,''
[arXiv:2207.09485 [hep-ph]].
%1 citations counted in INSPIRE as of 13 Aug 2022

%\cite{Alok:2022ovy}
\bibitem{Alok:2022ovy}
A.~K.~Alok, N.~R.~S.~Chundawat, A.~Mandal and T.~Sarkar,
%``Can neutron star discriminate between Dirac and Majorana neutrinos?,''
[arXiv:2208.02239 [hep-ph]].
%0 citations counted in INSPIRE as of 13 Aug 2022

%\cite{Murase:2015ndr}
\bibitem{Murase:2015ndr}
K.~Murase,
``Active Galactic Nuclei as High-Energy Neutrino Sources,''
%doi:10.1142/9789814759410\_0002
[arXiv:1511.01590 [astro-ph.HE]].
%56 citations counted in INSPIRE as of 12 Aug 2022

%\cite{Schuster:2001xi}
\bibitem{Schuster:2001xi}
C.~Schuster, M.~Pohl and R.~Schlickeiser,
``Neutrinos from active galactic nuclei as a diagnostic tool,''
Astron. Astrophys. \textbf{382}, 829 (2002)
%doi:10.1051/0004-6361:20011670
[arXiv:astro-ph/0111545 [astro-ph]].
%22 citations counted in INSPIRE as of 12 Aug 2022

%\cite{Giommi:2021bar}
\bibitem{Giommi:2021bar}
P.~Giommi and P.~Padovani,
``Astrophysical Neutrinos and Blazars,''
Universe \textbf{7}, no.12, 492 (2021)
%doi:10.3390/universe7120492
[arXiv:2112.06232 [astro-ph.HE]].
%7 citations counted in INSPIRE as of 12 Aug 2022


%\cite{Oikonomou:2019pmg}
\bibitem{Oikonomou:2019pmg}
F.~Oikonomou, K.~Murase and M.~Petropoulou,
``High-Energy Neutrinos from Blazar Flares and Implications of TXS 0506+056,''
EPJ Web Conf. \textbf{210}, 03006 (2019)
%doi:10.1051/epjconf/201921003006
[arXiv:1903.02006 [astro-ph.HE]].
%6 citations counted in INSPIRE as of 12 Aug 2022





\bibitem{PhysRevC.69.045803}
A.~M.~S.~Santos and D.~P.~Menezes
Phys. Rev. C \textbf{69}, 045803 (2004)  
  
%\cite{Mueller:1996pm}
\bibitem{Mueller:1996pm}
H.~Mueller and B.~D.~Serot,
``Relativistic mean field theory and the high density nuclear equation of state,''
Nucl. Phys. A \textbf{606}, 508-537 (1996)
%doi:10.1016/0375-9474(96)00187-X
[arXiv:nucl-th/9603037 [nucl-th]].
%388 citations counted in INSPIRE as of 12 Aug 2022  

%\cite{Glendenning:1991es}
\bibitem{Glendenning:1991es}
N.~K.~Glendenning and S.~A.~Moszkowski,
``Reconciliation of neutron star masses and binding of the lambda in hypernuclei,''
Phys. Rev. Lett. \textbf{67}, 2414-2417 (1991)
%doi:10.1103/PhysRevLett.67.2414
%715 citations counted in INSPIRE as of 12 Aug 2022


%\cite{Chatterjee:2021wsr}
\bibitem{Chatterjee:2021wsr}
D.~Chatterjee, J.~Novak and M.~Oertel,
``Structure of ultra-magnetised neutron stars,''
Eur. Phys. J. A \textbf{57}, no.8, 249 (2021)
%doi:10.1140/epja/s10050-021-00525-5
[arXiv:2108.13733 [nucl-th]].
%1 citations counted in INSPIRE as of 12 Aug 2022

\bibitem{1992ApJ...395..250G}
P.~Goldreich and A.~Reisenegger
The Astro. Journal, \textbf{395}, 250-258 (1992) 

%\cite{Sengupta:1998yk}
\bibitem{Sengupta:1998yk}
S.~Sengupta,
``Evolution of crustal magnetic fields in isolated neutron stars : combined effects of cooling and curvature of space-time,''
Astrophys. J. \textbf{501}, 792 (1998)
%doi:10.1086/305826
[arXiv:astro-ph/9801299 [astro-ph]].
%6 citations counted in INSPIRE as of 12 Aug 2022

\bibitem{2011MNRAS.413.2021G}
K.~Glampedakis, D.~I.~ Jones and L.~Samuelsson
Mon. Not. R. Astron. Soc. \textbf{413}, 2021-2030, (2011)
[arXiv:1010.1153v3[astro-ph]]


\bibitem{2012MNRAS.421.2722K},
Y.~Kojima and S.~Kisaka
Mon. Not. R. Astron. Soc. \textbf{421}, 2722-2730, (2012)
[arXiv:1201.1346[astro-ph.HE]]


%\cite{Passamonti:2016nmf}
\bibitem{Passamonti:2016nmf}
A.~Passamonti, T.~Akg\"un, J.~A.~Pons and J.~A.~Miralles,
``The relevance of ambipolar diffusion for neutron star evolution,''
Mon. Not. Roy. Astron. Soc. \textbf{465}, no.3, 3416-3428 (2017)
%doi:10.1093/mnras/stw2936
[arXiv:1608.00001 [astro-ph.HE]].
%32 citations counted in INSPIRE as of 12 Aug 2022
        
%\cite{Watanabe:2022cuv}
\bibitem{Watanabe:2022cuv}
C.~Watanabe, N.~Yoshinaga and S.~Ebata,
``Equations of State for Hadronic Matter and Mass-Radius Relations of Neutron Stars with Strong Magnetic Fields,''
Universe \textbf{8}, no.1, 48 (2022)
%doi:10.3390/universe8010048
%0 citations counted in INSPIRE as of 12 Aug 2022        

\bibitem{2015SSRv..191..111F}
L.~Ferrario, D.~de Martino and B.~T.~G{\"a}nsicke
Solar and Stellar Astrophysics, \textbf{191}, 1-4 (2015)
[arXiv:1504.08072[astro-ph.SR]]

%\cite{Drewes:2021fjx}
\bibitem{Drewes:2021fjx}
M.~Drewes, J.~McDonald, L.~Sablon and E.~Vitagliano,
``Neutrino Emissivities as a Probe of the Internal Magnetic Fields of White Dwarfs,''
Astrophys. J. \textbf{934}, 99 (2022)
%doi:10.3847/1538-4357/ac7874
[arXiv:2109.06158 [astro-ph.SR]].
%3 citations counted in INSPIRE as of 12 Aug 2022






\end{thebibliography}
\end{document}